\documentclass[two column]{aastex631}

\usepackage{subfigure}
\usepackage{amsmath}  
\usepackage{graphicx}
\usepackage{array}
\usepackage{hyperref}
\usepackage{multirow}
\usepackage{tablefootnote}
\usepackage{makecell}

\shorttitle{The rotation curve of M31}
\shortauthors{Wu et al.}

\begin{document}
\title{Mass Distribution of M31 from its Rotation Curve up to 167 kpc} 
\author[0009-0006-7193-4443]{Hao Wu}
\affiliation{Department of Astronomy, School of Physics, Peking University, Beijing 100871, China}
\affiliation{Kavli Institute for Astronomy and Astrophysics, Peking University, Beijing 100871, China}

\author[0000-0003-3250-2876]{Yang Huang}
\affiliation{School of Astronomy and Space Science, University of Chinese Academy of Science, Beijing 100049, China}
\affiliation{National Astronomical Observatories, Chinese Academy of Sciences, Beijing 100101, China}
\author[0000-0002-7727-1699]{Huawei Zhang}
\affiliation{Department of Astronomy, School of Physics, Peking University, Beijing 100871, China}
\affiliation{Kavli Institute for Astronomy and Astrophysics, Peking University, Beijing 100871, China}
\author{Qikang Feng}
\affiliation{Department of Astronomy, School of Physics, Peking University, Beijing 100871, China}
\affiliation{Kavli Institute for Astronomy and Astrophysics, Peking University, Beijing 100871, China}
\author{Cheng Liu}
\affiliation{Beijing Planetarium, Beijing Academy of Science and Technology, Beijing, 100044, China}
\received{30-Jun-2026}
\revised{09-Sep-2026}
\accepted{21-Sep-2026}

\correspondingauthor{Yang Huang: huangyang@ucas.ac.cn;\\Huawei Zhang: zhanghw@pku.edu.cn}
\begin{abstract}

The mass distribution of the Andromeda galaxy (M31) is fundamental for understanding its dark matter content, assembly history, and dynamical role within the Local Group. However, estimates of M31's total mass remain quite uncertain, partly because of limited dynamical constraints at large galactocentric radii. Moreover, M31's recent merger history has produced abundant tidal substructures and may have left its outer halo out of dynamical equilibrium, complicating equilibrium-based mass estimates.
We use over 8,500 tracers with line-of-sight velocities to constrain the rotation curve (RC) of M31 over $R=2$--167 kpc. In the disk, we derive the RC from a clean tracer sample using the Jeans equation and find good agreement with previous H~\textsc{I} measurements. In the halo, we use Gaussian mixture modeling to account for tidal substructures and Galactic foreground contamination and constrain the kinematics of the residual dynamically hot halo.
Assuming this halo population can be treated as an equilibrium tracer, we infer the outer RC and constrain M31's mass distribution for different velocity anisotropies $\beta$ and dark matter halo profiles. The NFW and Dekel--Zhao (DZ) models yield broadly consistent total masses ranging from $1.12^{+0.11}_{-0.10}\times10^{12}\,M_\odot$ to $1.63^{+0.21}_{-0.17}\times10^{12}\,M_\odot$ for $\beta=0.3$--0.7. In contrast, the Einasto profile favors substantially lower masses of $0.41_{-0.06}^{+0.09}$ to $0.86_{-0.16}^{+0.21}\times10^{12}\,M_\odot$, highlighting significant dependence on the adopted halo profile, although Einasto solutions are more in tension with independent mass estimates based on M31 satellite galaxies than the NFW and DZ models at fixed $\beta$. Overall, our analysis provides a comprehensive measurement of M31's mass distribution under the assumption of dynamical equilibrium, while also quantifying the systematic uncertainties associated with halo velocity anisotropy and the choice of dark matter density profile.

\end{abstract}


\section{Introduction}\label{sec:introduction}
The Andromeda galaxy (M31) is the closest giant spiral galaxy to the Milky Way (MW). As we reside within the disk of our MW, obtaining a global view of our own galaxy is inherently challenging. In contrast, M31 offers a unique external perspective, making it an ideal laboratory and critical benchmark for studying the hierarchical galaxy formation and evolution.

Over the past decades, deep photometric and spectroscopic surveys, including the Pan-Andromeda Archaeological Survey (PAndAS; \citealt{McConnachie2009,McConnachie2018}), the Panchromatic Hubble Andromeda Treasury (PHAT; \citealt{Dalcanton2012}), the Spectroscopic and Photometric Landscape of Andromeda's Stellar Halo (SPLASH; \citealt{Gilbert2009}), the Dark Energy Spectroscopic Instrument (DESI; \citealt{Dey2023}), and the Large Sky Area Multi-Object Fiber Spectroscopic Telescope (LAMOST; \citealt{Cui2012}), have revealed abundant evidence that M31 has undergone substantial dynamical perturbations in its recent history. Stellar density maps from PAndAS reveal extensive and highly asymmetric substructures in M31's halo, most notably the Giant Stellar Stream (GSS; \citealt{Ibata2001}) and the Northeastern and Western Shelves \citep{Ferguson2002,Fardal2007}. Complementary DESI observations have further identified numerous kinematically cold components in the inner halo \citep{Dey2023}. M31 also possesses a substantially thickened and dynamically heated stellar disk, with a steep age--velocity dispersion relation and an episode of enhanced star formation approximately $2$--$4\,{\rm Gyr}$ ago \citep{Dorman2015,Williams2015,Williams2017,Bhattacharya2019,Dalcanton2023}. Taken together, these disk and halo signatures have motivated major merger scenarios in which M31 experienced a substantial merger approximately 2–3 Gyr ago \citep{Hammer2018,Souza2018,Tsakonas2025}.

Determining the mass distribution of M31 is therefore essential for reconstructing its assembly history and interpreting the dynamics of its tidal debris and satellite population. Soon after M31 was recognized as an external galaxy, attempts were made to infer its mass distribution from its rotation curve (RC; e.g., \citealt{Babcock1939,Wyse1942}), providing early indications that its observed kinematics could not be explained by visible matter alone. Subsequent improved observations established that the RC remains approximately flat across the outer disk, providing evidence for an extended dark matter component \citep{Rubin1970}. With the establishment of the $\Lambda$CDM paradigm and its prediction of dark matter halos \citep{Ostriker1973}, substantial effort has been devoted to determining the total mass of M31. Existing estimates can be broadly grouped into five categories:

\begin{enumerate}
\item \textit{RC-based mass modeling}, in which M31's observed RC is used to constrain parametrized models of its underlying mass distribution \citep[e.g.,][]{Klypin2002,Carignan2006,Geehan2006,Seigar2008,Chemin2009,Corbelli2010,Tamm2012,Sofue2015,Zhang2024};

\item \textit{Halo-tracer dynamical modeling}, which uses the spatial distribution and velocities of various halo tracers, such as globular clusters and satellite galaxies, to constrain the gravitational potential through Jeans modeling, distribution-function fitting, or related mass estimators \citep[e.g.,][]{Evans&Wilkinson2000,Cote2000,Evans2003,Galleti2006,Lee2008,Watkins2010,Tollerud2012,Veljanoski2014,Hayashi2014};

\item \textit{Tidal stream and substructure modeling}, in which non-equilibrium substructures such as the GSS are modeled to constrain M31’s gravitational potential \citep[e.g.,][]{Ibata2004,Fardal2006,Fardal2013,Escala2022,Dey2023};

\item \textit{Local Group dynamical constraints}, which use the dynamics of the MW--M31 system and surrounding Local Group galaxies to constrain the mass of M31 \citep[e.g.,][]{vdm2012a,Penarrubia2014,Diaz2014,Penarrubia2016};

\item \textit{Cosmological simulations}, in which M31 analogues are selected according to observational constraints \citep[e.g.,][]{Zhai2020,Villanueva2023,Patel2023,Wempe2024}.
\end{enumerate}

These estimates typically span $\sim1$--$3\times10^{12}\,M_{\odot}$, representing a nearly threefold spread and highlighting the substantial systematic uncertainties in current determinations of M31's total mass.
This broad spread partly reflects the limited dynamical information available at large galactocentric radii. Although the H~\textsc{I} RCs derived by \citet{Chemin2009} and \citet{Corbelli2010} provide a robust characterization of the disk dynamics, they extend only to $\sim38$ kpc, well within the virial radius implied by most mass estimates. Estimates of the virial mass must therefore rely on more distant halo tracers. Previous studies have generally been limited to a few hundred globular clusters or satellites with highly incomplete phase-space information. In addition, the presence of abundant tidal debris and kinematically cold substructures in M31's halo can introduce biases into dynamical mass estimates if these components are treated as part of a single equilibrium halo population.

Beyond these observational limitations, \citet{Hammer2025} raised a more fundamental concern regarding the applicability of equilibrium-based mass estimators to M31. In their major merger model, the gas in the outer disk had not completed enough orbits since the merger to re-establish dynamical equilibrium. This model yielded a total dynamical mass of only $M_{\rm tot}(<R_{200})=4.5\times10^{11}\,M_{\odot}$. This substantially lower value raises a broader question about the applicability of equilibrium-based mass estimates to a recently disturbed system such as M31.

Recent spectroscopic surveys of M31, particularly SPLASH and DESI, have substantially increased the number of halo tracers with reliable LOS velocities to several thousand, while also enabling more effective identification and separation of Galactic foreground contaminants and kinematically cold tidal structures \citep[e.g.,][]{Gilbert2018,Dey2023}. This is particularly important because simulations have shown that phase-space-correlated debris can significantly bias the inferred tracer-density and velocity profiles, thereby introducing substantial systematic uncertainties into dynamical mass estimates \citep[e.g.,][]{Kafle2018b,Wang2018,Deason2021}. Identifying and removing such substructures therefore helps reduce an important source of bias in the mass inference. Nevertheless, the remaining dynamically hot halo cannot necessarily be assumed to be fully relaxed or in dynamical equilibrium. We therefore explicitly adopt dynamical equilibrium as a working assumption of our analysis, and the resulting circular velocity and mass profiles should be understood as being derived within this framework.

In this work, we use more than 8,500 tracers with measured LOS velocities to constrain the circular velocity profile of M31 over $R=2$--$167\,{\rm kpc}$. We derive the disk RC from disk tracers and constrain the outer halo using the LOS velocity dispersion of the dynamically hot halo after accounting for Galactic foreground and tidal substructures. Within the equilibrium assumption, we then constrain the mass distribution of M31 through parametrized mass modeling. Section~\ref{sec:Data} describes the data and sample selection. Sections~\ref{sec:Disk RC} and~\ref{sec:Halo RC} present the construction of the RC in the disk and halo regions, respectively. Section~\ref{sec:mass distribution} presents the mass modeling and results, and Section~\ref{sec:Summary} summarizes our conclusions.

\section{Sample}
\label{sec:Data}

\subsection{Sample in the Disk Region}
\label{subsec:disk sample}
At the distance of M31 (784 kpc; \citealt{Stanek1998}), spectroscopic studies are largely limited to intrinsically bright tracers, including globular clusters (GCs), supergiant stars, planetary nebulae (PNe), H~\textsc{ii} regions, and red giant branch (RGB) stars. For the disk component, we focus on supergiant stars, PNe, and H~\textsc{ii} regions, which are closely associated with the disk and are less susceptible to contamination from Galactic foreground stars and M31 halo populations.
The disk tracers used in this work are compiled from the following sources:
\begin{enumerate}

\item \textbf{Supergiant stars.} We compile red supergiants from \citet{Massey2009} and \citet{Massey2016}, yellow supergiants from \citet{Drout2009}, and the LAMOST supergiant catalog of \citet{Wu2025a}, yielding a final supergiant sample of 567 objects.

\item \textbf{Planetary nebulae.} We combine PNe catalogs from \citet{Merrett2006}, \citet{Halliday2006}, \citet{Sanders2012}, and \citet{Martin2018}. Only sources classified as M31 PNe are retained, while extended objects, non-PN contaminants, and PNe associated with M31 satellites or tidal substructures identified by \citet{Merrett2006} are excluded. The resulting sample contains 3,144 PNe.

\item \textbf{H~{\normalfont\textsc{ii}} regions.} We adopt H~\textsc{ii} regions from \citet{Sanders2012} and \citet{Chen2025}. The latter study identified emission-line objects using LAMOST spectra via a random forest classifier. 
In total, the final sample contains 372 H~\textsc{ii} regions.
\end{enumerate}

\begin{figure*}
\centering
    \begin{minipage}{0.49\textwidth}
    \centering
    \includegraphics[width=\textwidth]{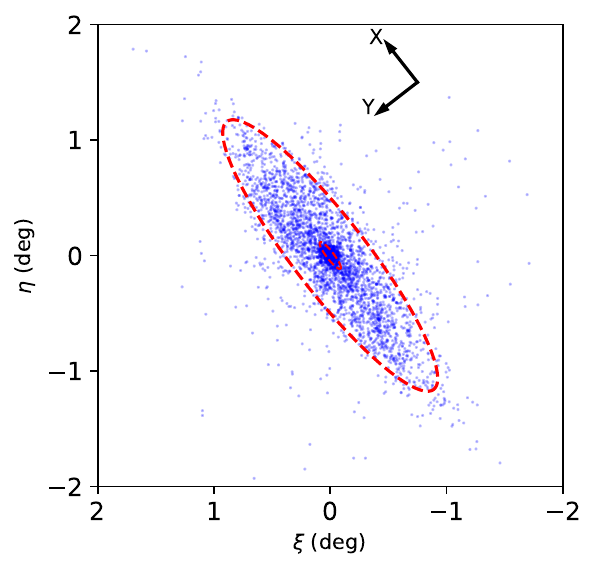}
    \end{minipage}
    \begin{minipage}{0.49\textwidth}
    \centering
    \includegraphics[width=\textwidth]{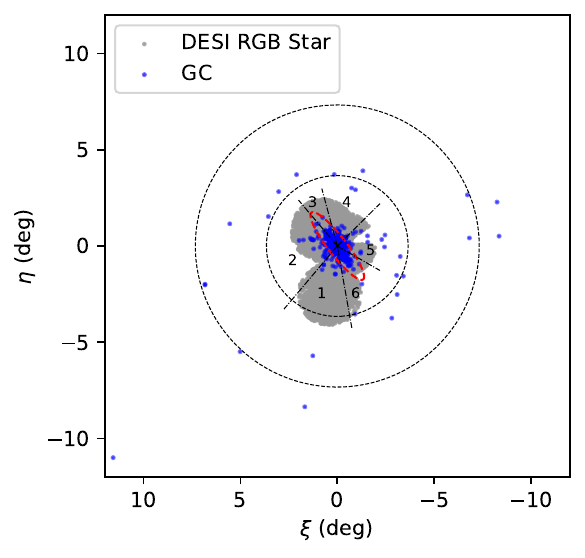}
    \end{minipage}
\caption{\textit{Left:} Spatial distribution of the 3,905 disk tracers used to construct the rotation curve (RC) of M31, shown in the M31-centric reference frame. The tracers are represented by blue dots. The inner and outer red dashed ellipses mark the projected disk radii of $R=2$ and $20\,{\rm kpc}$, respectively. The adopted $X$ and $Y$ axes are illustrated in the upper-right corner.
\textit{Right:} Spatial distribution of the tracers used to construct the RC in the halo region, shown in the same M31-centric reference frame. Gray dots represent all 7,348 stars from the DESI EDR, while blue dots denote the 512 globular clusters (GCs) used in this work. The six dash-dotted lines divide the DESI sample into six spatial regions (Regions 1–6), following the definitions of \citet{Dey2023}. The red dashed ellipse marks the projection of the disk-plane radius $R=30\,{\rm kpc}$. The inner and outer black dashed circles correspond to projected radii of 50 and 100 kpc, respectively.}
\label{fig:RC_sample}
\end{figure*}

To ensure data quality and homogeneity, we apply several additional processing steps to the compiled disk sample. We first cross-match sources among different catalogs using a matching radius of 1\arcsec.5. For objects with multiple observations, we retain the LOS velocity ($V_{\rm LOS}$) measurement with the smallest reported uncertainty.
The compiled sample combines observations obtained with several spectroscopic facilities, including Hectospec/MMT, WYFFOS/WHT, SITELLE/CFHT, LAMOST, and the Planetary Nebula Spectrograph (PNS) on the WHT. Systematic differences in the velocity zero-point may therefore exist among different catalogs constructed from different facilities \citep[e.g.,][]{Wu2025a}.
For each catalog, we calculate the $V_{\rm LOS}$ zero-point (ZPT) offset using sources cross-matched with the MMT-based reference catalog, as summarized in Table~\ref{tab:zpt_disk}. All measurements are then transformed onto the MMT-based velocity scale. For the catalog of \citet{Martin2018}, which has only a limited overlap with the MMT observations, we adopt the ZPT offset of $+1.8\,{\rm km\,s^{-1}}$ relative to the \citet{Halliday2006} catalog reported by \citet{Martin2018} and convert it to the MMT scale accordingly.
Finally, to further improve sample purity and remove potential foreground contamination, we cross-match all sources with \textit{Gaia} DR3 \citep{Gaiadr3} using a matching radius of 1\arcsec.5 and apply additional proper motion (PM) criteria to objects with available astrometric measurements \citep{Salomon2021}. Specifically, we retain only sources satisfying $|\mu_{\alpha*}| < 0.19~{\rm mas~yr^{-1}} + 3\sigma_{\mu_{\alpha*}}$ and $|\mu_\delta| < 0.19~{\rm mas~yr^{-1}} + 3\sigma_{\mu_\delta}$,
where $\sigma_{\mu_{\alpha*}}$ and $\sigma_{\mu_\delta}$ denote the PM uncertainties in right ascension and declination, respectively. Sources without PM measurements are retained, because the lack of reliable PMs is expected for many faint or distant M31 sources and should not be used as a criterion for rejection.
After applying these procedures, the final disk sample contains 442 supergiants, 3,127 PNe, and 336 H~\textsc{ii} regions, yielding a total of 3,905 disk tracers with measured LOS velocities. The spatial distribution of these tracers in the M31-centric frame is shown in the left panel of Fig.~\ref{fig:RC_sample}.

\begin{table}
\centering
\caption{Line-of-sight velocity zero-point offsets for the M31 disk-tracer catalogs, measured relative to the Hectospec/MMT velocity scale.}
\label{tab:zpt_disk}
\resizebox{1\linewidth}{!}{
\begin{tabular}{lcc}
\hline
Catalog & Spectrograph/Telescope & Zero-point offset (km\,s$^{-1}$) \\
\hline
\citet{Massey2009}   & Hectospec/MMT  & Reference \\
\citet{Drout2009}    & Hectospec/MMT  & Reference \\
\citet{Massey2016}   & Hectospec/MMT  & Reference \\
\citet{Merrett2006}  & PNS/WHT        & $-1.0$ \\
\citet{Halliday2006} & WYFFOS/WHT     & $-1.5$ \\
\citet{Martin2018}   & SITELLE/CFHT   & $+0.3$ \\
\citet{Wu2025a}      & LAMOST         & $-5.0$ \\
\citet{Chen2025}     & LAMOST         & $-5.0$ \\
\hline
\end{tabular}
}
\end{table}

\subsection{Sample in the Halo Region}
\label{subsec:halo sample}
Our halo sample consists of two components: stars from the DESI Early Data Release (DESI EDR; \citealt{Dey2023}) and GCs compiled from the literature.
The DESI EDR observations of M31 provide reliable LOS velocities and stellar parameters for more than 7,000 stars brighter than $m_z=21.5$. These observations primarily target (tip) RGB stars in the stellar halo of M31 and extend to projected radii ($R_p$) of approximately 55 kpc.
The DESI sample is strongly affected by abundant kinematically cold structures in M31's inner halo, as well as by residual contamination from Galactic foreground stars. This problem is further exacerbated by the DESI target selection, which is biased toward redder stars and therefore preferentially samples the metal-rich substructure populations that dominate much of the inner halo \citep{Dey2023}. Our goal is therefore to statistically separate these identifiable contaminants from the broader dynamically hot halo population, thereby reducing an important source of bias in the inferred velocity dispersion. The detailed mixture modeling is presented in Section~\ref{sec:Halo RC}.

The GC sample is compiled from the Revised Bologna Catalog Version 5 (RBCV5; \citealt{Galleti2004,Galleti2009}), from which we retain objects classified as ``GC'' or ``GC candidate'', together with the catalogs of \citet{Veljanoski2014}, \citet{Chen2016}, \citet{Caldwell2016}, and \citet{Sakari2016}. To construct a homogeneous GC sample, we apply several additional processing steps.
We first cross-match sources among different catalogs using a matching radius of 3\arcsec.0, appropriate for the extended nature of GCs. For objects with multiple measurements, we retain $V_{\rm LOS}$ with the smallest reported uncertainty. We further exclude sources with velocity uncertainties larger than $25\,{\rm km\,s^{-1}}$.
The compiled sample combines observations obtained with several different spectroscopic facilities and therefore may again suffer from velocity ZPT offsets. 
Similar to the disk sample, we calculate the $V_{\rm LOS}$ ZPT offsets for each catalog using sources cross-matched with the MMT-based reference catalog, as summarized in Table~\ref{tab:zpt_halo}. All measurements are then transformed onto the same MMT-based velocity scale adopted for the disk sample.

To minimize contamination from identifiable tidal substructures, we exclude any GC classified as a possible substructure member by either \citet{Veljanoski2014} or \citet{Mackey2019}, retaining only those classified as non-substructure members in both studies. This conservative selection removes GCs associated with known coherent structures and yields a final sample of 512 GCs. The spatial distributions of the DESI halo stars and the selected GCs are shown in the right panel of Fig.~\ref{fig:RC_sample}.

We also make use of the halo RGB star kinematic measurements from the SPLASH survey. Although the underlying stellar catalog is not publicly available, \citet{Gilbert2018} published the LOS velocity dispersion profile of the dynamically hot halo component out to $R\sim167$ kpc based on more than 2,000 spectroscopically observed M31 RGB stars. Using a Gaussian mixture model, they decomposed the observed LOS velocity distribution into multiple kinematic components and directly inferred the velocity dispersion of the dynamically hot halo population. As this approach closely matches our goal of reducing contamination from identifiable tidal substructures, we incorporate these measurements as an independent constraint on the kinematics of the residual dynamically hot component in Section~\ref{sec:Halo RC}.

Taken together, these selections and statistical decompositions reduce contamination from identifiable coherent substructures and Galactic foreground stars, providing cleaner constraints on the kinematics of the residual dynamically hot halo. In the subsequent mass modeling, this residual stellar halo component and the selected GC population are treated as dynamical tracers under the equilibrium assumption stated in Section~\ref{sec:introduction}.

\begin{table}
\centering
\caption{Line-of-sight velocity zero-point offsets for M31 globular cluster catalogs relative to the Hectospec/MMT velocity scale.}
\label{tab:zpt_halo}
\resizebox{\linewidth}{!}{
\begin{tabular}{lcc}
\hline
Catalog & Spectrograph/Telescope & Zero-point offset (km\,s$^{-1}$) \\
\hline
\citet{Caldwell2016}     & Hectospec/MMT   & Reference \\
RBCV5                    & --              & $+2.0$ \\
\citet{Veljanoski2014}   & ISIS/WHT        & $-0.4$ \\
\citet{Chen2016}         & LAMOST          & $-5.0$ \\
\citet{Sakari2016}       & APOGEE          & $-1.6$ \\
\hline
\end{tabular}}
\end{table}

\subsection{Velocity Transformation}
\label{subsec:Velocity Trans}

Throughout this work, positions and velocities are expressed in an M31-centric reference frame. We adopt the following basic parameters for M31: central coordinates (RA, Dec) $=(10.68^\circ,41.27^\circ)$ \citep{Evans2010}, inclination angle $i=77.5^\circ$ \citep{Walterbos1988,deVaucouleurs1991}, position angle ${\rm PA}=37.5^\circ$ \citep{Chemin2009,Corbelli2010}, and distance $d=784$ kpc \citep{Stanek1998}.

Because our tracers span more than $20^\circ$ on the sky, the LOS projections of both the solar reflex motion and the bulk motion of M31 vary significantly across the sample. It is therefore necessary to apply a position-dependent correction to each source before constructing the RC.

To correct for the solar reflex motion, we adopt
$(U_{\odot},V_{\odot},W_{\odot})=(7.01,10.13,4.95)\,{\rm km\,s^{-1}}$ \citep{Huang2015} and a circular speed at the solar position of
$V_{\rm c}(R_0)=234.04\,{\rm km\,s^{-1}}$ \citep{Zhou2023}. The heliocentric LOS velocity, $V_{\rm LOS}$, is transformed into the Galactocentric frame as
\begin{align}
    V_{\rm Gal} =& \,V_{\rm LOS} + 244.17 \sin{l}\cos{b} \notag \\
    &+7.01 \cos{l}\cos{b}+4.95\sin{b}\text{,}
\end{align}
where $l$ and $b$ are the Galactic longitude and latitude of the tracer.

We then transform the velocities to the M31-centric rest frame by subtracting the LOS projection of M31's bulk motion at the position of each tracer \citep{vdm2008}:
\begin{equation}
V_{\rm pec} = V_{\rm Gal} - V_{\rm rad}^{\rm M31}\cos\rho - V_{\rm tran}^{\rm M31}\sin\rho \cos(\theta_{\rm PA}-\theta_{\rm tran}^{\rm M31}) ,
\end{equation}
where $V_{\rm rad}^{\rm M31}=-109.0\,{\rm km\,s^{-1}}$ is the Galactocentric radial velocity of M31 \citep{vdm2008,vdm2012a,Salomon2016}, $V_{\rm tran}^{\rm M31}=46.7\,{\rm km\,s^{-1}}$ is the Galactocentric transverse velocity of M31 \citep{Wu2025b}, and $\theta_{\rm tran}^{\rm M31}=92.0^\circ$ is the position angle of the transverse motion vector of M31 \citep{Wu2025b}. Here $\rho$ is the angular separation between the tracer and the center of M31, and $\theta_{\rm PA}$ is the position angle of the tracer relative to the center of M31. Both position angles are measured from north through east.

The adopted solar motion and M31 transverse motion parameters differ slightly from those used in previous studies \citep[e.g.,][]{Veljanoski2014,Gilbert2018,Zhang2024}, primarily because we adopt the latest available measurements. The resulting differences in $V_{\rm pec}$ are typically only $2$--$3\,{\rm km\,s^{-1}}$, well below the typical velocity uncertainties.
In the halo region, where the outer RC is inferred from the LOS velocity dispersion (Section~\ref{sec:Halo RC}), these parameter choices have a negligible impact because the resulting changes are far smaller than the intrinsic velocity dispersion of the halo tracers.

\section{Rotation Curve in the Disk Region}
\label{sec:Disk RC}
\subsection{Kinematic Model}
\label{subsec:disk rc model}

We construct the disk RC using the disk sample selected in Section~\ref{subsec:disk sample}. The sky coordinates of all disk tracers are transformed into projected major-axis ($X$) and minor-axis ($Y$) coordinates in the M31-centric frame. The $X$ and $Y$ axes point toward the northeast and southeast directions, respectively, as illustrated in the top-right corner of the left panel of Fig.~\ref{fig:RC_sample}.
We then establish a cylindrical coordinate system $(R,\phi,Z)$ centered on M31, with the $R$--$\phi$ plane aligned with the disk plane. In this system, $R$ increases radially outward from the center, and $\phi$ is measured from M31's projected major axis and increases counterclockwise.
Assuming that all disk tracers lie in the disk plane and neglecting their vertical terms, the galactocentric radius and azimuthal angle are given by $R=\sqrt{X^2+(Y/\cos i)^2}$ and $\phi = \arctan\left(\frac{Y/\cos i}{X}\right)$, respectively, where $i=77.5^\circ$.

To minimize contamination from bulge and halo populations, we retain only tracers within $2<R<20$ kpc, corresponding to the region enclosed by the red dashed lines in the left panel of Fig.~\ref{fig:RC_sample}. 
The outer limit is chosen conservatively to focus on the main disk region and to avoid larger radii where halo contamination and tidal substructures become increasingly important \citep{Zhang2024}. This is further supported by the radial velocity dispersion profile shown below, which generally declines over this radial range, indicating that the retained tracers are consistent with a disk-dominated population. This selection leaves 2,940 disk tracers. 

Assuming an axisymmetric gravitational potential of M31, the circular velocity profile can be written as:
\begin{equation}\label{eq:grav_disk}
V_{\text{c}}^{2}(R) = R\frac{\partial \Phi}{\partial R}.
\end{equation}
The Jeans equation in cylindrical coordinates is given by \citep{Binney2008}:
\begin{equation}
\frac{\partial(\nu \langle V_{R}^{2}\rangle)}{\partial R} + \frac{\partial(\nu \langle V_{R}V_{Z}\rangle)}{\partial Z} + \nu\left(\frac{\langle V_{R}^{2}\rangle - \langle V_{\phi}^{2}\rangle}{R} + \frac{\partial \Phi}{\partial R}\right) = 0,
\end{equation}
where $\nu$ is the spatial density distribution of the disk tracers. We neglect the cross-term $\langle V_RV_Z\rangle$ and its vertical gradient, which are expected to be small compared to the other terms. By substituting the above equation into Eq.~\ref{eq:grav_disk}, we obtain
\begin{equation}\label{eq:jeans}
V_{\text{c}}^{2}(R) = \langle V_{\phi}^{2}\rangle - \langle V_{R}^{2}\rangle\left(1+\frac{\partial \text{ln}\nu}{\partial\text{ln}R} + \frac{\partial \text{ln}\langle V_{R}^{2}\rangle}{\partial\text{ln}R}\right) .
\end{equation}
 
We therefore need to determine the radial profiles of $\langle V_R^2\rangle$, $\langle V_\phi^2\rangle$, and $\nu(R)$ that enter Eq.~\ref{eq:jeans}.
We begin by decomposing the observed peculiar LOS velocity as:
\begin{equation}\label{eq:vel_decompose}
V_{\text{pec}} = V_{R}\sin{\phi}\sin{{i}} + V_{\phi}\cos{\phi}\sin{{i}} + V_Z\cos{i},
\end{equation}
where $V_R$, $V_\phi$ and $V_Z$ are the radial, azimuthal and vertical velocity components defined in the cylindrical coordinate system $(R,\phi,Z)$. For the highly inclined disk of M31, the vertical contribution $V_Z\cos i$ is negligible to $V_{\rm pec}$.
We then exploit this projection geometry to estimate the radial and azimuthal second-moment profiles entering Eq.~\ref{eq:jeans}. For tracers with very small $|\cos\phi|$, the azimuthal component contributes negligibly to $V_{\rm pec}$, allowing the radial velocity to be estimated as $V_R=V_{\rm pec}/(\sin\phi\sin i)$. This allows us to measure the radial profile of $\langle V_{R}^{2}\rangle$. Conversely, for tracers with large $|\cos\phi|$ (i.e., $|\sin\phi|\approx0$), their azimuthal velocities can be approximated as $V_\phi=V_{\rm pec}/(\cos\phi\sin i)$, which provide a favorable reference for the inferred $\langle V_\phi^2\rangle$ profile.

We first select 162 sources with $|\cos\phi|\leq 0.06$, where the LOS velocity is dominated by the radial component and thus allows a reliable estimate of $V_R$. This subsample is used exclusively to determine the radial velocity dispersion profile. The mean radial velocity of these sources is close to zero ($0.9\pm92.0\,{\rm km\,s^{-1}}$). We therefore adopt $\langle V_R\rangle=0$\,km\,s$^{-1}$, such that $\langle V_R^2\rangle$ can be interpreted as the radial velocity dispersion $\sigma_R^2$.

\begin{figure}
    \centering
    \begin{minipage}{0.9\columnwidth}
        \centering
        \includegraphics[width=\textwidth]{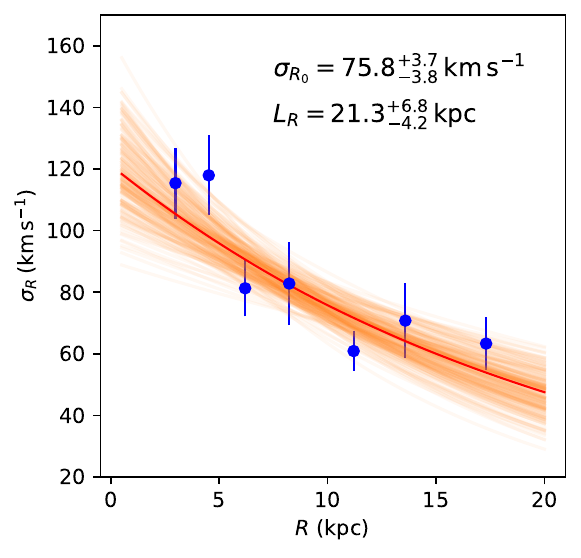}
    \end{minipage}
    \vspace{0.3cm}
    \begin{minipage}{0.9\columnwidth}
        \centering
        \includegraphics[width=\textwidth]{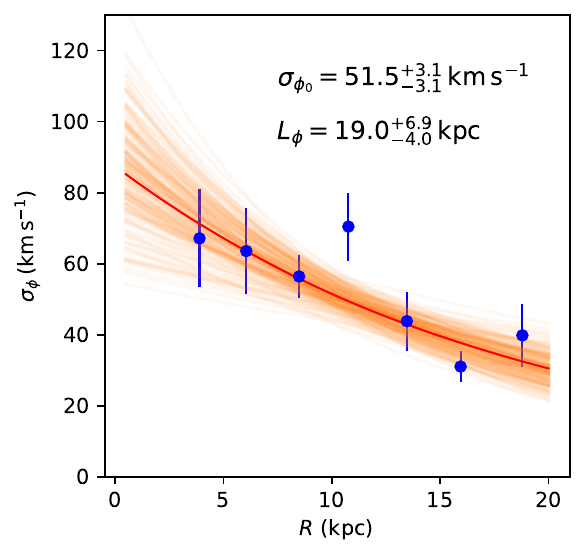}
    \end{minipage}
    \caption{
    Radial profiles of the disk velocity dispersions. \textit{Top:} Radial velocity dispersion, $\sigma_R$, for sources with $|\cos \phi| \leq 0.06$. \textit{Bottom:} Azimuthal velocity dispersion, $\sigma_\phi$, for sources with $|\cos \phi| \geq 0.997$. Blue symbols show the measured velocity dispersions and their uncertainties in different radial bins. The red curves show the best-fitting profiles, while the orange curves show posterior draws from the model parameter distributions to illustrate the associated uncertainties.}
    \label{fig:disk_sigma}
\end{figure}

To measure the radial profile of $\sigma_R$, we divide the sample into seven radial bins, adopting a minimum bin width of 1.5 kpc and enlarging the bin width when necessary until each bin contains more than 20 sources. For each bin, we perform 500 bootstrap resamplings to estimate $\sigma_R$ and its uncertainty. The resulting measurements are shown as the blue symbols in the top panel of Fig.~\ref{fig:disk_sigma}. 
Following \citet{Huang2016} and \citet{Zhang2024}, we model the radial velocity dispersion as an exponentially declining function of radius:
\begin{equation}
\sigma_R(R) = \sigma_{R_0} \exp ( -\frac{R-R_0}{L_{R}}),
\label{eq:disk sigma_R}
\end{equation}
where $R_0=10$ kpc is a fixed reference radius \citep{Zhang2024}, and $\sigma_{R_0}$ denotes the radial velocity dispersion at that radius. Using Markov Chain Monte Carlo (MCMC) fitting, we obtain $\sigma_{R_0}=75.8^{+3.7}_{-3.8}\,{\rm km\,s^{-1}}$ and $L_R=21.3^{+6.8}_{-4.2}$ kpc. The best-fitting profile is shown as the red curve in the top panel of Fig.~\ref{fig:disk_sigma}.

The second moment of the azimuthal velocity in Eq.~\ref{eq:jeans} can be written as
$\langle V_{\phi}^{2}\rangle=\langle V_{\phi}\rangle^{2}+\sigma_{\phi}^{2}$,
where $\sigma_{\phi}$ is the azimuthal velocity dispersion. We estimate $\sigma_\phi$ using sources for which the LOS velocity is dominated by the azimuthal component. Specifically, we select 119 near-major-axis sources with $|\cos\phi|\ge0.997$ (i.e., $|\sin\phi|\le0.075$). For these sources, the contribution of $V_R$ to the LOS velocity is negligible, so $V_\phi$ can be robustly estimated as
$V_\phi=V_{\rm pec}/(\cos\phi\sin i)$.
We again divide these sources into seven radial bins, adopting a minimum bin width of 2 kpc and enlarging the bin width when necessary until each bin contains more than 10 sources. In each bin, we compute both the mean azimuthal velocity and the azimuthal velocity dispersion. The uncertainty in velocity dispersion is estimated from 500 bootstrap resamples, each constructed by drawing $N$ sources with replacement from a bin of $N$ sources. The measured dispersions are shown as the blue symbols in the bottom panel of Fig.~\ref{fig:disk_sigma}. The jump at $R\sim11$ kpc may be associated with a local substructure, such as the ring \citep{Gordon2006, Lewis2015, Williams2017}, or with a bar-induced effect \citep{Blana2018}. As for $\sigma_R$, we model the $\sigma_\phi$ profile as
\begin{equation}
\sigma_\phi(R)=\sigma_{\phi_0}\exp\left(-\frac{R-R_0}{L_{\phi}}\right),
\label{eq:disk sigma_phi}
\end{equation}
with $R_0=10\,{\rm kpc}$. The MCMC fit gives
$\sigma_{\phi_0}=51.5^{+3.1}_{-3.1}\,{\rm km\,s^{-1}}$
and $L_{\phi}=19.0^{+6.9}_{-4.0}\,{\rm kpc}$. The best-fitting profile is shown as the red curve in the bottom panel of Fig.~\ref{fig:disk_sigma}.

\begin{figure}
	\includegraphics[width=8cm,height=6cm]{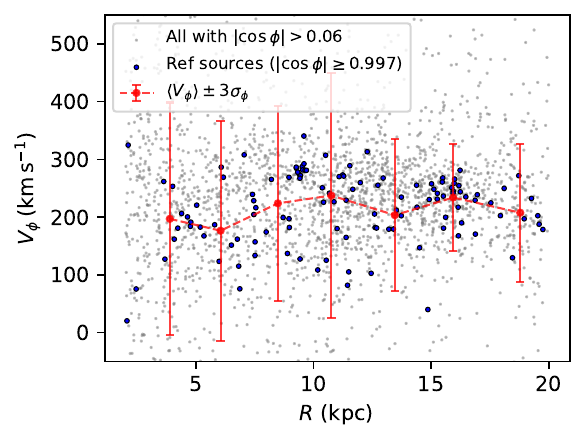}
    \caption{Selection of sources used to construct the disk rotation curve (RC). Gray points show the proxy azimuthal velocities, $V_{\rm pec}/(\cos\phi\sin i)$, for all 2,778 sources with $|\cos\phi|>0.06$. Blue points show the $V_\phi$ values of the near-major-axis reference subsample with $|\cos\phi|\geq0.997$. Red symbols mark the mean $V_\phi$ of this reference subsample in seven radial bins, with error bars indicating the corresponding $3\sigma_\phi$ range measured in each bin. Sources lying within this reference envelope are selected for the subsequent disk RC analysis.}
    \label{fig:disk_vphi}
\end{figure}

We next estimate $\langle V_\phi\rangle$ from the broader sample of 2,778 sources with $|\cos\phi|>0.06$. For individual objects in this sample, the radial term in Eq.~\ref{eq:vel_decompose} may not be negligible. However, under the assumption $\langle V_R\rangle=0\,{\rm km\,s^{-1}}$, the radial contribution is expected to average out within each radial bin if the bin contains a sufficiently large and azimuthally representative sample. In this statistical sense, the mean azimuthal velocity in each radial bin can be approximated as
\begin{equation}\label{eq:Vphi_bin}
\langle V_\phi\rangle_{\rm bin} \simeq \frac{1}{N_{\rm bin}}
\sum_{j=1}^{N_{\rm bin}} \frac{V_{{\rm pec},j}}{\cos\phi_j\sin i},
\end{equation}
where $j$ indexes the sources in the bin and $N_{\rm bin}$ is the number of sources in that bin.
This approximation may nevertheless be affected by non-circular or peculiar motions, especially because some disk tracers may be associated with spiral arms or local kinematic structures. We therefore use the near-major-axis measurements described above to define an empirical rotation envelope. The $V_\phi$ values of the 119 near-major-axis sources are shown as blue points in Fig.~\ref{fig:disk_vphi}. For comparison, the proxy values $V_{\rm pec}/(\cos\phi\sin i)$ for all 2,778 sources are shown as gray points.

In each of the seven reference bins, we define the rotation envelope as the range within $3\sigma_\phi$ of the mean $V_\phi$ measured from the near-major-axis sources. We use this envelope to identify sources whose inferred $V_\phi$ values are consistent with the expected disk rotation at a given radius. Among the 2,778 sources with $|\cos\phi|>0.06$, 1,983 objects (71.4\%) fall within this envelope. We retain these sources for the subsequent analysis, reducing the impact of objects whose LOS velocities may be strongly affected by radial motions, spiral-arm streaming, or other local peculiar motions.

\subsection{Results}\label{subsec:disk rc result}

By assuming that the M31 disk follows an exponential surface density profile with a scale length $L_d$ and substituting Eq.~\ref{eq:disk sigma_R} into Eq.~\ref{eq:jeans}, the circular velocity can be written as
\begin{equation}\label{eq:jeans transform}
\begin{aligned}
V_{\rm c}^{2}(R)
&=
\langle V_{\phi}^{2}\rangle
-
\sigma_{R}^{2}
\left(1-\frac{R}{L_d}-\frac{2R}{L_R}\right) \\
&=
\langle V_{\phi}\rangle^{2}
+
\sigma_{\phi}^{2}
-
\sigma_{R}^{2}
\left(1-\frac{R}{L_d}-\frac{2R}{L_R}\right).
\end{aligned}
\end{equation}

\begin{figure}
\centering
	\includegraphics[width=\columnwidth]{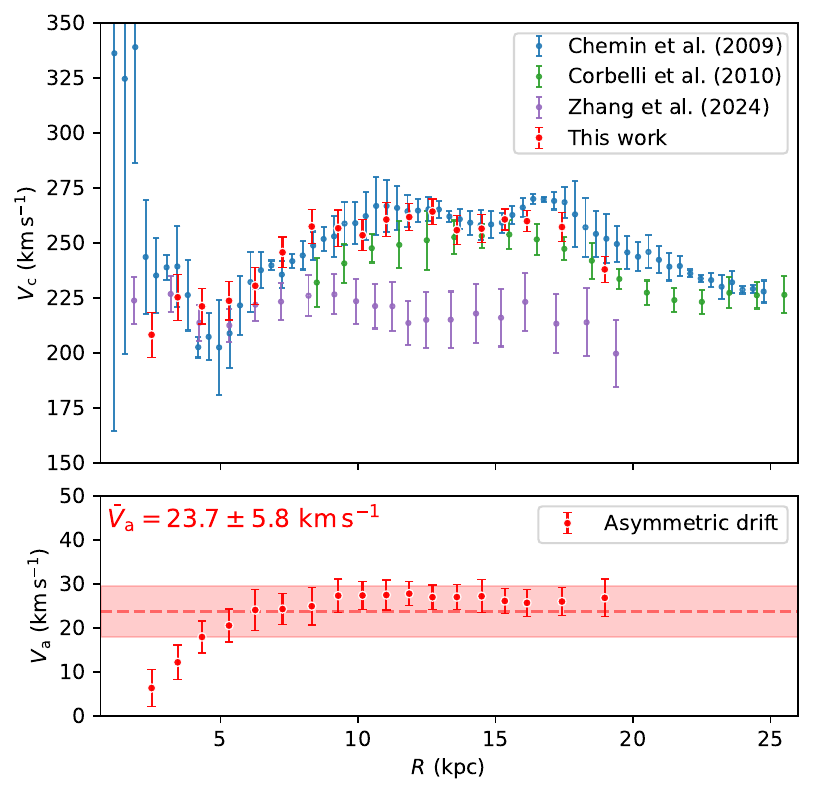}
    \caption{M31 rotation curve (RC) in the disk region ($2\,\mathrm{kpc} < R < 20\,\mathrm{kpc}$). The upper panel presents a comparison between our newly derived RC and those from previous studies. Our results are shown as red symbols, while the H~\textsc{I}-based RC measurements from \citet{Chemin2009} and \citet{Corbelli2010} are indicated by blue and green symbols, respectively. The more recent measurements by \citet{Zhang2024} are represented by purple symbols. The bottom panel shows the asymmetric drift $V_{\rm a}$ in relation to radius derived in this work, with the red dashed line and the shaded area indicating the mean $V_{\rm a}$ and 1$\sigma$ scatter.}
    \label{fig:disk_rc}
\end{figure}

We then divide the 1,983 selected sources into 18 radial bins, adopting a minimum bin width of 0.85 kpc and enlarging the bin width when necessary until each bin contains more than 100 sources for statistical robustness.
We estimate $V_{\rm c}$ and its uncertainty using a bootstrap Monte Carlo (MC) procedure based on Eq.~\ref{eq:jeans transform}, which accounts for finite-sample fluctuations and the dominant uncertainties in the quantities entering this equation. For each radial bin, we generate 500 realizations. In each realization, for a radial bin containing $N$ sources, we randomly draw $N$ sources with replacement from that bin, following the standard bootstrap resampling scheme. For each selected source, we randomly sample its azimuthal velocity from a Gaussian distribution centered on the inferred $V_\phi$, with a standard deviation given by the uncertainty in $V_\phi$ propagated from $V_{\rm LOS}$. We then recalculate the mean azimuthal velocity, $\langle V_\phi\rangle$, from these sampled velocities using Eq.~\ref{eq:Vphi_bin}.
For other related terms in Eq.~\ref{eq:jeans transform}, we propagate the model uncertainties through MC sampling. In each realization, we sample one set of parameters for the fitted $\sigma_R$ and $\sigma_\phi$ profiles from their posterior distributions derived in Section~\ref{subsec:disk rc model}. The sampled $\sigma_R$ parameters include $L_R$, whose median value is $L_R=21.3_{-4.2}^{+6.8}\,{\rm kpc}$. The disk scale length is independently sampled from a Gaussian distribution, $L_d\sim\mathcal{N}(5.3,0.5^2)\,{\rm kpc}$ \citep{Courteau2011}. For each realization, we evaluate $\sigma_R$, $\sigma_\phi$, and the correction term in Eq.~\ref{eq:jeans transform} at the mean radius of the sources in the corresponding bin. We then compute $V_{\rm c}^2$ from Eq.~\ref{eq:jeans transform} and assign $V_{\rm c}=\sqrt{V_{\rm c}^2}$.
In each radial bin, we adopt the median of the resulting $V_{\rm c}$ distribution as the circular velocity, and estimate its uncertainty as half of the 16th--84th percentile range. The final circular velocity profile is listed in Table~\ref{tab:RC_disk} and shown in the upper panel of Fig.~\ref{fig:disk_rc}.

The asymmetric drift in each radial bin, defined as
$V_{\rm a}=V_{\rm c}-\langle V_{\phi}\rangle$, is shown in the bottom panel of Fig.~\ref{fig:disk_rc}. The mean value is $23.7\pm5.8\,{\rm km\,s^{-1}}$. This level of asymmetric drift is consistent with the composition of our tracer sample, which is dominated by old-population PNe (1,253 out of 1,983 sources; 63.2\%), with the remaining sources consisting of younger supergiant stars and H~\textsc{ii} regions. Since older stellar populations in M31 generally exhibit larger asymmetric drift \citep{Quirk2019}, a moderate asymmetric drift is expected for this mixed but predominantly old tracer sample.

Compared with previous measurements, our circular velocity profile shows excellent agreement with the H~\textsc{I}-based RCs of \citet{Chemin2009} and \citet{Corbelli2010}. In particular, over $4$--$15\,{\rm kpc}$, our results closely follow those of \citet{Chemin2009}. At larger radii, the profile gradually declines and lies between the two H~\textsc{I} RCs. We also detect a noticeable dip near $4\,{\rm kpc}$, which may be associated with bar-induced perturbations \citep{Chemin2009,Liu2025}. In contrast, our circular velocity profile differs from that reported by \citet{Zhang2024}. This difference may reflect an incomplete correction for asymmetric drift in their analysis, as well as possible contamination of their sample by Galactic foreground stars.

\begin{table}
	\centering
	\caption{The rotation curve of M31's disk region.}
	\label{tab:RC_disk}
	\begin{tabular}{ccc}
	\hline
	$R$ (kpc) & $V_{\rm c}\,\rm(km\,s^{-1})$ & $\sigma_{V_{\rm c}}\,\rm(km\,s^{-1})$ \\
	\hline
	2.49 & 208.24 & 10.23 \\
	3.44 & 225.29 & 10.46 \\
	4.32 & 221.15 & 8.06 \\
	  5.30 & 223.69 & 8.68 \\
	6.26 & 230.50 & 8.39 \\
	7.26 & 246.69 & 7.00 \\
	8.31 & 258.44 & 7.75 \\
	9.28 & 256.66 & 8.30 \\
	10.16 & 253.49 & 7.01 \\
	11.03 & 260.62 & 7.60 \\
	11.86 & 261.75 & 6.12 \\
	12.70 & 264.23 & 5.71 \\
	13.59 & 255.79 & 6.53 \\
	14.49 & 256.47 & 6.39 \\
	15.33 & 260.68 & 4.78 \\
	16.14 & 259.89 & 4.73 \\
	17.41 & 257.23 & 6.69 \\
    18.97 & 237.95 & 5.80 \\
	\hline
	\end{tabular}
\end{table}

\section{Rotation Curve in the Halo Region}
\label{sec:Halo RC}
\subsection{Kinematic Model}\label{subsec:halo rc model}
We now construct the kinematic model to constrain the RC in the halo region. From the halo sample selected in Section~\ref{subsec:halo sample}, we retain only objects with cylindrical radius $R>30\,{\rm kpc}$, where $R$ is computed in the same way as in the disk analysis. This cut, indicated by the red dashed curve in the right panel of Fig.~\ref{fig:RC_sample}, is applied to reduce contamination from the M31 disk population. The resulting halo sample consists of 92 GCs and 5,430 DESI stars.

For these halo tracers, the true three-dimensional (3D) spherical distance from the center of M31, $r$, cannot be directly measured. Instead, each tracer has only a projected two-dimensional radius,
\begin{equation}
R_p = d\sqrt{\eta^2+\xi^2},
\end{equation}
where $d=784\,{\rm kpc}$ is the adopted distance to M31 \citep{Stanek1998}, and $\eta$ and $\xi$ are M31-centric coordinates. The only available kinematic observable is the peculiar LOS velocity $V_{\rm pec}$.

Because full 3D positions and velocities are unavailable for the halo tracers, we adopt a deprojection approach similar to that used by \citet{Zhang2024}. To connect the kinematics of the residual dynamically hot halo to the gravitational potential, we assume that this component can be approximated as a steady-state tracer population with a constant velocity anisotropy parameter $\beta$. Under these assumptions, the spherical Jeans equation is
\begin{equation}
\frac{\mathrm{d} (\rho \sigma_r^2)}{\mathrm{d} r}+2\frac{\beta}{r}\rho\sigma_r^2=-\rho\frac{\mathrm{d}\Phi}{\mathrm{d}r},
\label{eq:halo jeans eq}
\end{equation}
where $\rho$ is the tracer number density, $\sigma_r$ is the radial velocity dispersion, and $\Phi$ is the total gravitational potential. The velocity anisotropy is $\beta = 1 - \frac{\sigma_{\theta}^2 + \sigma_{\phi}^2}{2\sigma_r^2}$, where $\sigma_{\theta}$ and $\sigma_{\phi}$ are the tangential velocity dispersions in spherical coordinates. The potential is related to the circular velocity by
\begin{equation}
V_{\rm c}^2(r)=r\frac{\mathrm{d}\Phi}{\mathrm{d}r}.
\label{eq:halo potential eq}
\end{equation}
We assume that the halo tracer density follows a power-law profile,
$\rho=\rho_0 r^{\alpha_{\rm 3D}}$ \citep{Courteau2011,Ibata2014},
and that the radial velocity dispersion follows
$\sigma_r\propto r^\gamma$ \citep{Veljanoski2014,Gilbert2018,Zhang2024}. Substituting these relations into Eqs.~\ref{eq:halo jeans eq} and \ref{eq:halo potential eq} gives:
\begin{equation}
V_{\rm c}^2 = -\sigma_r^2(2\gamma+\alpha_{\rm 3D}+2\beta).
\label{eq:halo trans}
\end{equation}

We then connect the intrinsic halo kinematics to the observable LOS velocity using the deprojection equation for a spherical system \citep{Binney1982,Mamon2005,Mamon2010},
\begin{equation}\label{eq:halo deproj}
\sigma_{\rm LOS}^2(R_p) = \frac{2}{\Sigma (R_p)}
\int_{R_p}^{\infty} \left(1-\beta\frac{R_p^2}{r^2}\right)
\rho(r)\sigma_r^2 \frac{r\,\mathrm{d}r}{\sqrt{r^2-R_p^2}} .
\end{equation}
This expression assumes spherical symmetry and equal angular velocity dispersions, $\sigma_\theta=\sigma_\phi$. Here $\sigma_{\rm LOS}$ denotes the peculiar LOS velocity dispersion\footnote{Hereafter, the LOS velocity dispersion refers to the peculiar LOS velocity dispersion, calculated after removing the solar reflex motion and the bulk motion of M31, as described in Section~\ref{subsec:Velocity Trans}.}, and $\Sigma(R_p)$ is the projected tracer 2D surface density. We describe the latter with a power-law profile,
$\Sigma(R_p)=\Sigma_0 R_p^{\alpha_{\rm 2D}}$ \citep{Courteau2011,Ibata2014}.
Combining Eq.~\ref{eq:halo deproj} with the Jeans-based expression for the circular velocity in Eq.~\ref{eq:halo trans}, we obtain
\begin{equation}
\begin{aligned}
\sigma_{\rm LOS}^2(R_p)=&\frac{2\rho_0/\Sigma_0}{-2\gamma-\alpha_{\rm 3D}-2\beta}R_p^{-\alpha_{\rm 2D}}\\
&\times\int_{R_p}^{\infty}\left(1-\beta\frac{R_p^2}{r^2}\right)
r^{\alpha_{\rm 3D}+1}V_{\rm c}^2\frac{\mathrm{d}r}{\sqrt{r^2-R_p^2}} .
\end{aligned}
\label{eq:sigmaLOS_rp}
\end{equation}

This expression acts as a forward projection from the 3D circular velocity profile, $V_{\rm c}(r)$, to the observable 2D LOS dispersion profile, $\sigma_{\rm LOS}(R_p)$. Thus, within this framework, any assumed $V_{\rm c}(r)$ predicts a projected dispersion profile $\sigma_{\rm LOS}(R_{\rm p})$, which can be compared directly with the observations to constrain the halo circular velocity profile. This requires the measured $\sigma_{\rm LOS}$--$R_p$ relation, together with the density slopes $\alpha_{\rm 3D}$ and $\alpha_{\rm 2D}$, the normalization ratio $\rho_0/\Sigma_0$ at 1\,kpc, and the halo anisotropy parameter $\beta$.

\subsection{Halo Velocity Dispersion}\label{subsec:halo vel disp}

We first derive the $\sigma_{\rm LOS}$--$R_{\rm p}$ relation separately for the 5,430 DESI EDR stars and the 92 GCs in our final halo-tracer sample. As noted in Section~\ref{sec:Data}, the DESI stellar sample is strongly affected by tidal substructures and Galactic foreground contamination. We therefore adopt a Gaussian mixture model (GMM), following \citet{Gilbert2018}, to infer the LOS velocity dispersion of the broader dynamically hot halo component after statistically accounting for these contaminants. In this framework, the observed LOS velocity distribution is modeled as a superposition of a broad hot-halo component, kinematically cold components, and Galactic foreground stars. The dispersion of the hot component is inferred statistically without assigning each star to a single population.

\subsubsection{Gaussian Mixture Model for halo stars}

We first measure the $\sigma_{\rm LOS}$--$R_p$ relation for the 5,430 DESI stars using the GMM framework. Unlike the SPLASH survey, which consists of many isolated pencil-beam fields, the DESI observations provide broader and more contiguous coverage of M31's inner halo, where tidal debris is highly spatially structured. We therefore divide the DESI stellar sample into six spatial regions following the convention of \citet{Dey2023}, so that the GMM can account for region-dependent substructure components. The region boundaries are shown by the black dash-dotted lines in the right panel of Fig.~\ref{fig:RC_sample}. Regions 1--6 contain 1,914, 1,392, 345, 956, 440, and 383 halo tracers, respectively.
Region 1 is strongly dominated by the GSS \citep{Dey2023}, making a reliable Gaussian decomposition of the dynamically hot halo component difficult. We therefore exclude Region 1 from the stellar halo dispersion analysis, but use it in the following analysis to construct empirical priors for the foreground star components.

For the stellar sample outside Region 1, we estimate the halo $\sigma_{\rm LOS}$ in $R_p$ bins using a hierarchical Bayesian model. We jointly model Regions 2, 4, and 5, which contain enough stars to be divided into the same radial bins. Within each radial bin, the dynamically hot halo population\footnote{Hereafter, the halo component refers to the dynamically hot
component inferred by the GMM after accounting for identifiable kinematically cold structures and Galactic foreground contamination.} is assumed to share a single mean LOS velocity, $\mu_{\rm LOS}^{\rm Halo}$, and dispersion, $\sigma_{\rm LOS}^{\rm Halo}$, across these regions. The tidal substructure and foreground components are instead allowed to vary independently from region to region.
Regions 3 and 6 are excluded from the joint inference because they contain too few stars to be subdivided into the same radial bins as Regions 2, 4, and 5. We therefore model these two regions separately in the subsequent analysis.

For the halo stars in Regions 2, 4, and 5, we divide the combined sample into three projected radial bins:
$6.6\,{\rm kpc}\,(0.48^{\circ})<R_p\leq17.8\,{\rm kpc}\,(1.3^{\circ})$,
$17.8\,{\rm kpc}\,(1.3^{\circ})<R_p\leq27.4\,{\rm kpc}\,(2.0^{\circ})$, and
$27.4\,{\rm kpc}\,(2.0^{\circ})<R_p\leq39.6\,{\rm kpc}\,(2.9^{\circ})$.
These bins span the full projected radial range of the combined sample. In each radial bin, we include only spatial regions with more than 250 stars in the fit, so that the mixture components are sufficiently constrained and the inferred halo dispersion is statistically robust.

In each radial bin and spatial region, we model the stellar LOS velocity distribution as a mixture of three components: the dynamically hot halo, kinematically cold components (KCCs), and the Galactic foreground. Separate priors are assigned to each class. For the dynamically hot halo, we adopt the same prior on $\sigma_{\rm LOS}^{\rm Halo}$ as \citet{Gilbert2018}, together with a similar but slightly narrower prior on $\mu_{\rm LOS}^{\rm Halo}$. For the KCCs, the number of components and their velocity priors are chosen primarily from the substructure identifications and characterizations in Figs.~6--7 of \citet{Dey2023}.

Region 2 requires additional treatment because the overlap between the DESI and SPLASH footprints provides independent constraints on local substructures. Using SPLASH RGB stars, \citet{Escala2022} showed that sources in this region at $R_p>25\,{\rm kpc}$ with LOS velocities around $\sim -300\,{\rm km\,s^{-1}}$ ($V_{\rm pec}\sim0\,{\rm km\,s^{-1}}$) are predominantly associated with tidal debris rather than the dynamically hot halo, with a velocity dispersion of $45$--$55\,{\rm km\,s^{-1}}$. We therefore include a corresponding KCC in the outer two radial bins of Region 2, with priors on $\mu_{\rm LOS}$ and $\sigma_{\rm LOS}$ guided by \citet{Escala2022}.
To reduce the degeneracy between the halo and tidal-debris components, we impose upper limits on the halo fraction in Region 2: 30\% for $17.8\,{\rm kpc}<R_p\leq27.4\,{\rm kpc}$ and 60\% for $27.4\,{\rm kpc}<R_p\leq39.6\,{\rm kpc}$. With these constraints, the excess near $V_{\rm pec}\sim0\,{\rm km\,s^{-1}}$ is modeled primarily by the KCC component, consistent with the tidal-debris interpretation of \citet{Escala2022}.

For the foreground component, we construct a foreground-dominated subsample to guide the GMM priors. We select stars from Regions 1--6 with $V_{\rm pec}>151\,{\rm km\,s^{-1}}$ and $R_p>2.5^\circ$, where foreground contaminants are clearly separated from M31 members in velocity space (see Fig.~8 of \citealt{Dey2023}). Although Region 1 contains rich KCCs, the foreground population remains well isolated from M31 stars in this velocity range. The cut $V_{\rm pec}>151\,{\rm km\,s^{-1}}$, approximately corresponding to $V_{\rm LOS}>-150\,{\rm km\,s^{-1}}$, is also consistent with previous foreground selections in M31 studies \citep[e.g.,][]{Escala2020,Escala2022}, although it may exclude a small number of genuine M31 members.
From this foreground-dominated subsample, we obtain
$\mu_{\rm LOS}^{\rm Fgd}=263.6\,{\rm km\,s^{-1}}$ and
$\sigma_{\rm LOS}^{\rm Fgd}=26.5\,{\rm km\,s^{-1}}$, and estimate an overall foreground fraction of approximately 9\% for stars with $R_p>2.5^\circ$. These quantities are used as references for the foreground priors in the GMM. The full set of prior choices is listed in Table~\ref{tab:GMM_priors}.

\begin{deluxetable*}{ccccccccccc}
\tabletypesize{\footnotesize} %
\tablewidth{0pt}
\tablecaption{Results of Gaussian Mixture Modeling for Regions 2, 4, and 5. The parameters represent the peculiar LOS velocities, given as the median (50th percentile) of the posterior distributions, with uncertainties derived from the 16th and 84th percentiles. \label{tab:GMM_results}}
\tablehead{
    \colhead{$R_{p}$} & 
    \colhead{$\mu_{\rm LOS}^{\rm Halo}$} & 
    \colhead{$\sigma_{\rm LOS}^{\rm Halo}$} & 
    \colhead{Region} & 
    \colhead{$f_{\rm Halo}$} & 
    \colhead{$\mu_{\rm LOS}^{\rm KCC1}$} & 
    \colhead{$\sigma_{\rm LOS}^{\rm KCC1}$} & 
    \colhead{$f_{\rm KCC1}$} & 
    \colhead{$\mu_{\rm LOS}^{\rm KCC2}$} & 
    \colhead{$\sigma_{\rm LOS}^{\rm KCC2}$} & 
    \colhead{$f_{\rm KCC2}$} \\
    \colhead{(kpc)} & 
    \colhead{(km\,s$^{-1}$)} & 
    \colhead{(km\,s$^{-1}$)} & 
    \colhead{} & 
    \colhead{} & 
    \colhead{(km\,s$^{-1}$)} & 
    \colhead{(km\,s$^{-1}$)} & 
    \colhead{} & 
    \colhead{(km\,s$^{-1}$)} & 
    \colhead{(km\,s$^{-1}$)} & 
    \colhead{}
}
\startdata
\multirow{3}{*}{\makecell[c]{$6.6 < R_p \leq 17.8$}} & \multirow{3}{*}{$-15.9^{+5.0}_{-4.9}$} & \multirow{3}{*}{$108.0^{+8.2}_{-7.2}$} & 2 & $0.36^{+0.05}_{-0.05}$ & $30.6^{+10.7}_{-6.9}$ & $45.9^{+11.3}_{-8.4}$ & $0.32^{+0.09}_{-0.07}$ & $163.3^{+11.6}_{-8.7}$ & $52.2^{+10.2}_{-10.7}$  & $0.29^{+0.05}_{-0.07}$\\
\cline{4-11}
  & & & 4 &  $0.19^{+0.09}_{-0.11}$ & $69.1^{+6.5}_{-6.4}$ &  $71.1^{+4.9}_{-5.6}$  & $0.73^{+0.09}_{-0.08}$ & $-218.9^{+23.3}_{-16.9}$ & $53.4^{+15.8}_{-14.6}$  & $0.06^{+0.03}_{-0.02}$ \\
\cline{4-11}
  & & & 5 &  $0.66^{+0.12}_{-0.18}$ & $108.0^{+30.1}_{-26.7}$ & $74.2^{+4.3}_{-8.8}$  & $0.27^{+0.12}_{-0.09}$ & $-127.6^{+41.3}_{-32.6}$ & $46.1^{+23.8}_{-22.9}$  & $0.05^{+0.07}_{-0.04}$ \\
\hline
\multirow{2}{*}{\makecell[c]{$17.8 < R_p \leq 27.4$}} & \multirow{2}{*}{$-17.4^{+5.0}_{-4.9}$} & \multirow{2}{*}{$88.8^{+16.6}_{-9.9}$} & 2 & $0.16^{+0.09}_{-0.10}$ & $101.8^{+6.9}_{-7.4}$ & $38.8^{+6.6}_{-5.5}$ & $0.25^{+0.05}_{-0.04}$ & $-19.0^{+6.6}_{-6.1}$ & $59.5^{+3.6}_{-5.0}$  & $0.51^{+0.09}_{-0.09}$\\
\cline{4-11}
  & & & 4 &  $0.10^{+0.08}_{-0.07}$ & $87.6^{+8.9}_{-7.7}$ &  $73.5^{+4.3}_{-5.7}$  & $0.69^{+0.06}_{-0.07}$ & $-158.5^{+7.2}_{-6.5}$ & $32.9^{+5.8}_{-5.0}$  & $0.13^{+0.02}_{-0.02}$ \\
\hline
\makecell[c]{$27.4 < R_p \leq 39.6$} & $-11.1^{+4.4}_{-4.5}$ & $100.3^{+10.3}_{-8.3}$ & 2 & $0.53^{+0.05}_{-0.06}$ & $2.4^{+4.4}_{-4.6}$ & $34.4^{+5.2}_{-4.2}$ & $0.39^{+0.06}_{-0.05}$ & -- & --  & --\\
\enddata
\end{deluxetable*}

\begin{figure*}
\centering
    \begin{minipage}{\textwidth}
    \centering
    \includegraphics[width=\textwidth]{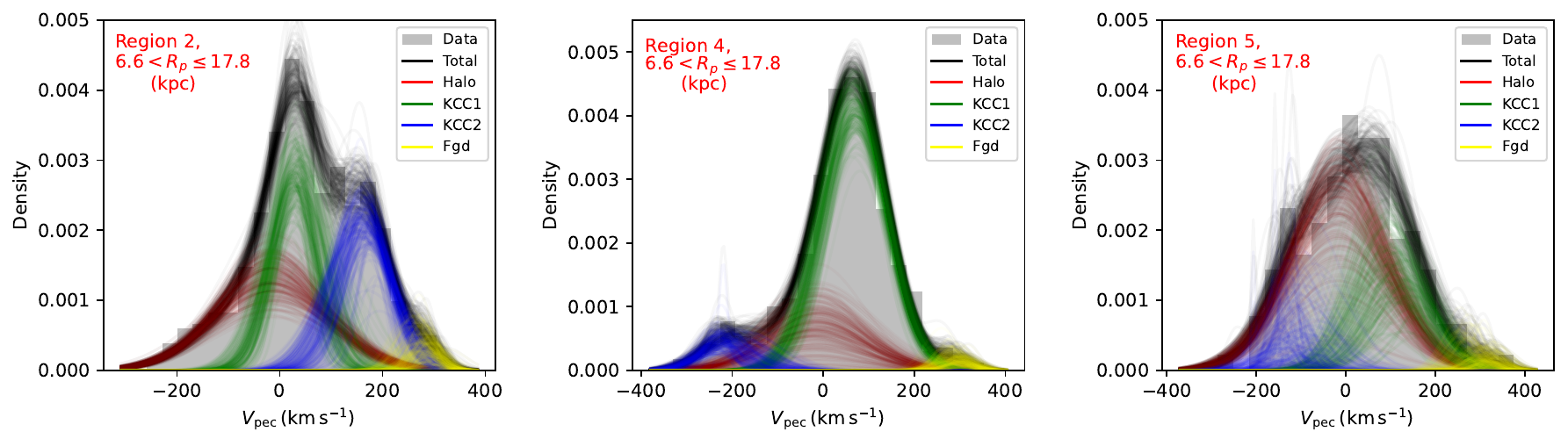}
    \end{minipage}
    \begin{minipage}{\textwidth}
    \centering
    \includegraphics[width=\textwidth]{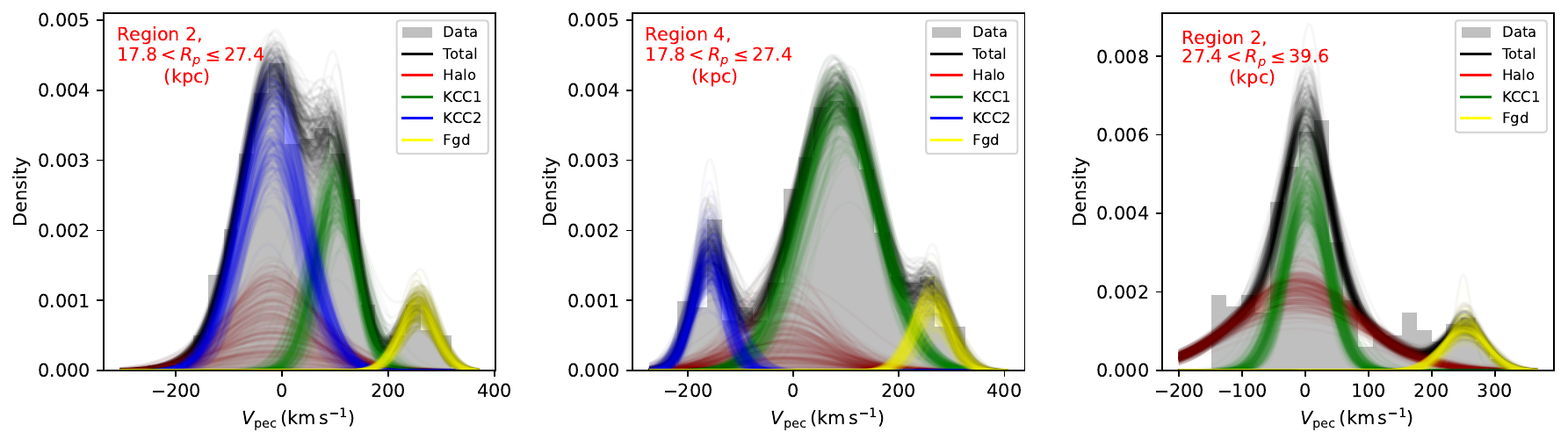}
    \end{minipage}
\caption{Gaussian Mixture Modeling results for Regions 2, 4 and 5 based on the posterior distributions of the model parameters, indicating the decomposition of the observed $V_{\rm pec}$ distributions into distinct kinematic components. Each panel is labeled with the corresponding region and $R_p$ range in the top-left corner. The gray histogram shows the density distribution of the observed $V_{\rm pec}$ data, overlaid with the total model (black curve) and individual Gaussian mixture components: Halo (red), KCC1 (green), KCC2 (blue), and foreground (Fgd; yellow). The colored curves represent 200 random posterior samples drawn from the NUTS chains, illustrating the uncertainty in each model component.}
\label{fig:GMM245}
\end{figure*}

Given the large number of free parameters in the hierarchical GMM, we sample the posterior distribution using \texttt{NumPyro}\footnote{\url{https://github.com/pyro-ppl/numpyro}} \citep{Bingham2019,phan2019}, a probabilistic programming framework built on \texttt{JAX}\footnote{\url{https://github.com/google/jax}} \citep{jax2018}. We use the No-U-Turn Sampler (NUTS; \citealt{Hoffman2014}), an adaptive Hamiltonian Monte Carlo algorithm \citep{Duane1987,Brooks2011} well suited to high-dimensional hierarchical models. In each radial bin, we run two chains with 4,000 warm-up iterations and 6,000 sampling iterations per chain. The derived GMM parameters are presented in Table~\ref{tab:GMM_results} and Fig.~\ref{fig:GMM245}.

\begin{figure*}
\centering
    \includegraphics[width=0.8\textwidth]{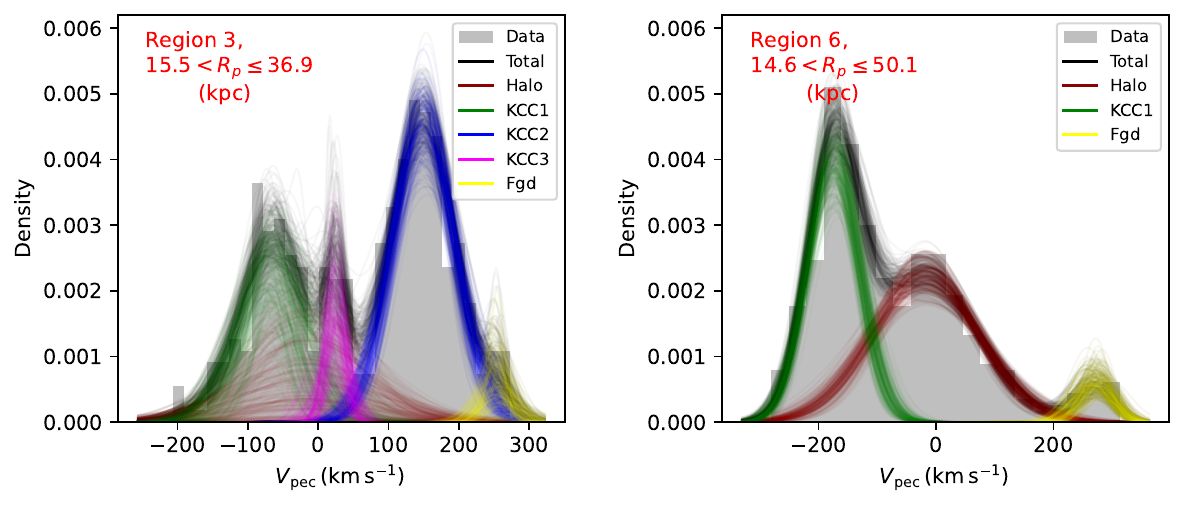}
\caption{Gaussian Mixture Modeling results for Regions 3 and 6 based on the posterior distributions of the model parameters, illustrating the decomposition of the observed $V_{\rm pec}$ distributions into distinct kinematic components. Each panel is labeled with the corresponding region and $R_p$ range in the top-left corner. The histograms and colored lines follow the same convention as in Fig.~\ref{fig:GMM245}.}
\label{fig:GMM36}
\end{figure*}

For the remaining two regions, Regions 3 and 6, we apply the same GMM-based decomposition to the LOS velocity distribution in each region separately. The model parameters and adopted priors are listed in Table~\ref{tab:GMM_priors_36}. The posteriors are again sampled with NUTS, and the results are summarized in Table~\ref{tab:GMM_results_region36} and shown in Fig.~\ref{fig:GMM36}.
Although Region 6 was not analyzed by \citet{Dey2023}, our GMM fit identifies a possible KCC-like component. This feature may be associated with tidal debris, or it may reflect contamination from M31 disk stars given its spatial location and mean velocity.

\begin{deluxetable*}{cccccccccccccc}
\tabletypesize{\scriptsize}
\setlength{\tabcolsep}{2pt} 
\renewcommand{\arraystretch}{1.5} 
\tablewidth{0pt}
\tablecaption{Results of Gaussian Mixture Modeling for Regions 3 and 6. The parameters represent the peculiar LOS velocities, given as the median (50th percentile) of the posterior distributions, with uncertainties derived from the 16th and 84th percentiles. \label{tab:GMM_results_region36}}
\tablehead{
    \colhead{$R_{p}$} & 
    \colhead{$\mu_{\rm LOS}^{\rm Halo}$} & 
    \colhead{$\sigma_{\rm LOS}^{\rm Halo}$} & 
    \colhead{Region} & 
    \colhead{$f_{\rm Halo}$} & 
    \colhead{$\mu_{\rm LOS}^{\rm KCC1}$} & 
    \colhead{$\sigma_{\rm LOS}^{\rm KCC1}$} & 
    \colhead{$f_{\rm KCC1}$} & 
    \colhead{$\mu_{\rm LOS}^{\rm KCC2}$} & 
    \colhead{$\sigma_{\rm LOS}^{\rm KCC2}$} & 
    \colhead{$f_{\rm KCC2}$} & 
    \colhead{$\mu_{\rm LOS}^{\rm KCC3}$} & 
    \colhead{$\sigma_{\rm LOS}^{\rm KCC3}$} & 
    \colhead{$f_{\rm KCC3}$} \\
    \colhead{(kpc)} & 
    \colhead{(km\,s$^{-1}$)} & 
    \colhead{(km\,s$^{-1}$)} & 
    \colhead{} & 
    \colhead{} & 
    \colhead{(km\,s$^{-1}$)} & 
    \colhead{(km\,s$^{-1}$)} & 
    \colhead{} & 
    \colhead{(km\,s$^{-1}$)} & 
    \colhead{(km\,s$^{-1}$)} & 
    \colhead{} & 
    \colhead{(km\,s$^{-1}$)} & 
    \colhead{(km\,s$^{-1}$)} & 
    \colhead{}
}
\startdata
\makecell[c]{$15.5 < R_p \leq 36.9$} & $-18.4_{-5.1}^{+5.1}$ & $93.8^{+34.2}_{-19.3}$ & 3 & $0.13^{+0.15}_{-0.10}$ & $-66.6^{+8.5}_{-7.7}$ & $44.6^{+10.0}_{-12.6}$ & $0.27^{+0.08}_{-0.10}$ & $149.2^{+4.8}_{-4.6}$ & $44.0^{+5.0}_{-4.4}$  & $0.51^{+0.04}_{-0.04}$ & $24.1^{+3.9}_{-7.6}$ & $14.8^{+6.9}_{-4.5}$  & $0.06^{+0.03}_{-0.03}$ \\
\hline
\makecell[c]{$14.6 < R_p \leq 50.1$} & $-17.9_{-4.7}^{+4.7}$ & $91.1^{+8.5}_{-7.9}$ & 6 & $0.51^{+0.04}_{-0.04}$ & $-174.7^{+4.7}_{-4.7}$ & $44.3^{+3.7}_{-3.3}$ & $0.44^{+0.04}_{-0.04}$ & -- & --  & -- & -- & --  & -- \\
\enddata
\end{deluxetable*}

Overall, the inferred fractions of dynamically hot halo stars are relatively low across different regions and $R_p$ ranges. This is expected because M31's inner halo is strongly dominated by tidal debris \citep{Ibata2014}, and because the DESI target selection preferentially includes metal-rich substructure populations \citep{Dey2023}. These low halo fractions highlight the importance of the GMM approach: without accounting for the kinematically cold structures and foreground stars, the inferred halo LOS velocity dispersion could be substantially biased.

\subsubsection{Parameterization of Halo Velocity Dispersion with Projected Radius}\label{subsubsec:Parameterization of Halo Velocity Dispersion}

\begin{figure*}
\centering
\includegraphics[width=0.47\textwidth]{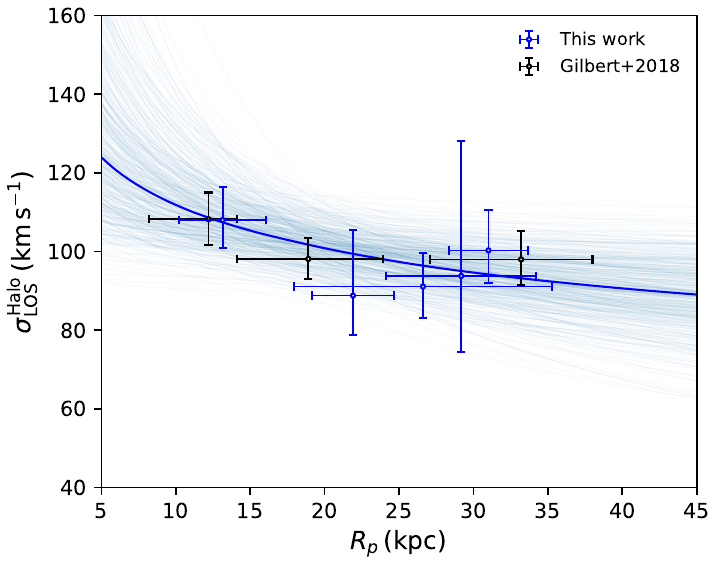}
\hfill
\includegraphics[width=0.47\textwidth]{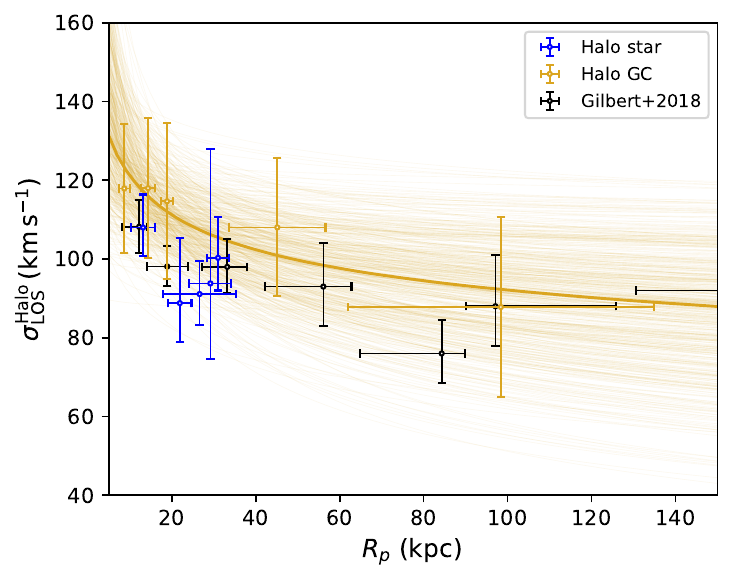}
\vspace{1em}
\includegraphics[width=0.7\textwidth]{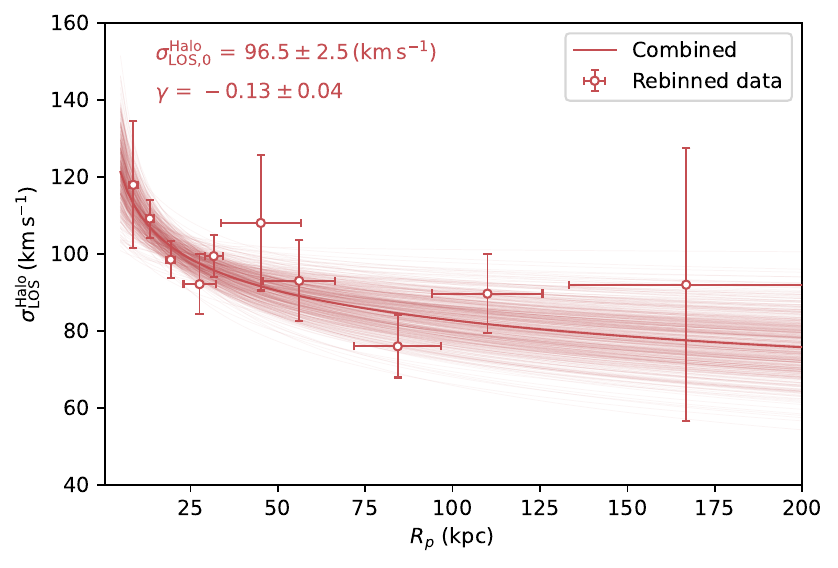}
\caption{Peculiar line-of-sight velocity dispersion ($\sigma_{\rm LOS}^{\rm Halo}$) of the M31 halo as a function of projected radius ($R_p$). $Top-left$ panel: $\sigma_{\rm LOS}^{\rm Halo}$ measurements for halo stars. Blue symbols show our new measurements based on DESI EDR halo stars, while black symbols show those from \citet{Gilbert2018}. The solid blue curve indicates the best-fit power-law model, with 500 posterior draws overplotted in light blue to illustrate the associated uncertainties. $Top-right$ panel: $\sigma_{\rm LOS}^{\rm Halo}$ for halo globular clusters (GCs) compared with halo stars. Yellow symbols show the velocity dispersion measurements of 92 selected halo GCs, grouped into five $R_p$ bins. The solid yellow curve indicates the best-fit power-law model for the GC data, with 500 random posterior draws shown as light yellow lines. For comparison, the DESI (blue) and SPLASH (black) halo star results are also shown. $Bottom$ panel: Combined fit to all halo tracers. The red points show the weighted mean dispersions in the ten rebinned $R_p$ bins. The solid red curve represents the best-fit power-law model to the rebinned data, with 500 random posterior draws plotted in light red to illustrate the associated uncertainties.}
\label{fig:halo_disp}
\end{figure*}

Using the derived $\sigma_{\rm LOS}^{\rm Halo}$ values at different $R_p$, we parameterize the halo LOS velocity dispersion profile as a power law, following \citet{Veljanoski2014}, \citet{Gilbert2018}, and \citet{Zhang2024}:
\begin{equation}
\sigma_{\rm LOS}^{\rm Halo}=\sigma_{\rm LOS,0}^{\rm Halo}\left(\frac{R_p}{R_{p,0}}\right)^\gamma ,
\label{eq:halo_los_dispersion}
\end{equation}
where the scale radius is fixed to $R_{p,0}=30\,{\rm kpc}$. We infer the posterior distributions of $\sigma_{\rm LOS,0}^{\rm Halo}$ and $\gamma$ using an MCMC fit, with the results shown in the top-left panel of Fig.~\ref{fig:halo_disp}. The best-fit parameters are
$\sigma_{\rm LOS,0}^{\rm Halo}=94.7\pm5.9\,{\rm km\,s^{-1}}$
and $\gamma=-0.15\pm0.11$. These values are in excellent agreement with the results of \citet{Gilbert2018}, who similarly derived a smooth, declining halo dispersion profile with $\sigma_{\rm LOS,0}^{\rm Halo}=96.6\pm3.2\,{\rm km\,s^{-1}}$ and $\gamma=-0.12\pm0.05$ using SPLASH RGB stars over $R_p=8$--$200\,{\rm kpc}$.

In addition to the halo stars, we examine the $\sigma_{\rm LOS}^{\rm Halo}$--$R_{\rm p}$ relation for the 92 halo GCs selected in Section~\ref{subsec:halo rc model}. For this sample, from which known substructure members were excluded, we directly estimate the LOS velocity dispersion in projected radial bins without applying a GMM decomposition. To ensure a reliable estimate of the velocity dispersion, we divide the GCs into five projected radial bins, requiring each bin to contain more than 15 sources for statistical robustness. In each bin, $\sigma_{\rm LOS}^{\rm Halo}$ and its uncertainty are estimated using 500 bootstrap resamples, each constructed by drawing $N$ sources with replacement from a bin of $N$ sources. The resulting GC LOS velocity dispersions are shown as yellow symbols in the top-right panel of Fig.~\ref{fig:halo_disp}. Fitting the same power-law model gives
$\sigma_{\rm LOS,0}^{\rm Halo}=105.9\pm9.0\,{\rm km\,s^{-1}}$
and $\gamma=-0.12\pm0.08$.
The GC dispersions are generally higher than those from the stellar sample, but they agree within the $1\sigma$ uncertainties. Together, the stellar and GC tracers are consistent with a mildly declining halo velocity dispersion profile.

We then construct a composite halo dispersion profile by combining the $\sigma_{\rm LOS}^{\rm Halo}$ measurements from DESI stars, SPLASH stars, and GCs. To avoid oversampling measurements at closely spaced projected radii, we group the data points into ten common $R_p$ bins. For each bin, the representative $R_p$ and $\sigma_{\rm LOS}^{\rm Halo}$ are computed as error-weighted averages of the measurements in that bin. The resulting rebinned points are separated by $\gtrsim 4\,{\rm kpc}$ in $R_p$ and extend to $R_p\sim167\,{\rm kpc}$, ensuring a more uniform radial sampling of the halo dispersion profile. The rebinned dispersion values are listed in Table~\ref{tab:halo_dispersion} and shown in the bottom panel of Fig.~\ref{fig:halo_disp}.
We note that the LOS velocity-dispersion profile shows an overall declining trend with increasing projected radius, although some local fluctuations are present. In particular, the two outermost measurements at $R_{\rm p}\simeq110$ and $167\,{\rm kpc}$ are higher than that at $R_{\rm p}\simeq84\,{\rm kpc}$. However, given the limited number of tracers at these large radii and the corresponding measurement uncertainties, these differences remain consistent within the uncertainties and do not constitute statistically significant evidence for an upturn in the dispersion profile. We therefore retain all ten rebinned measurements in our fiducial power-law fit, obtaining $\sigma_{\rm LOS,0}^{\rm Halo}=96.5\pm2.5\,{\rm km\,s^{-1}}$ and $\gamma=-0.13\pm0.04$. Nevertheless, the outermost measurements probe the largest radii, where the number of available tracers is smallest, orbital timescales are longest, and the assumption of dynamical equilibrium is correspondingly less secure. We therefore further assess their influence on the inferred mass distribution in Section~\ref{subsec:mass_results} by repeating the mass modeling after progressively excluding the outermost measurements.

Compared with our composite profile, \citet{Zhang2024} inferred systematically higher $\sigma_{\rm LOS}^{\rm Halo}$ values. This difference may arise because their analysis did not explicitly account for identifiable tidal substructures and Galactic foreground contamination when estimating the velocity dispersion of the broader dynamically hot halo component.
\begin{table}
	\centering
	\caption{Re-binned measurements of the M31 halo line-of-sight velocity dispersion ($\sigma_{\rm LOS}^{\rm Halo}$) in relation to the projected radius ($R_p$).}
	\label{tab:halo_dispersion}
	\begin{tabular}{cccc}
	\hline
	$R_p$ (kpc) & $\sigma_{R_p}$ (kpc) & $\sigma_{\rm LOS}^{\rm Halo}\,\rm(km\,s^{-1})$ & $\sigma_{\sigma_{\rm LOS}^{\rm Halo}}\,\rm(km\,s^{-1})$ \\ 
	\hline
	8.66 & 1.26 & 117.93 & 16.48 \\
    13.38 & 1.16 & 109.17 & 4.93 \\
	19.37 & 1.21 & 98.50 & 4.80 \\
    27.62 & 4.59 & 92.14 & 7.78 \\
	31.73 & 2.62 & 99.43 & 5.54 \\
    45.13 & 11.48 & 108.04 & 17.54 \\
	56.10 & 10.25 & 93.00 & 10.50 \\
	84.30 & 12.50 & 76.00 & 8.10 \\
	109.97 & 15.75 & 89.67 & 10.31 \\
	166.80 & 33.50 & 92.00 & 35.50 \\
	\hline
	\end{tabular}
\end{table}

\subsection{Parameter Selection}\label{subsec:halo parameter}

As described in Section~\ref{subsec:halo rc model}, we constrain the halo RC from the observed $\sigma_{\rm LOS}(R_p)$ profile using Eq.~\ref{eq:sigmaLOS_rp}. This calculation uses the value of $\gamma$ determined above and requires the 3D and 2D density slopes, $\alpha_{\rm 3D}$ and $\alpha_{\rm 2D}$, the normalization ratio $\rho_0/\Sigma_0$, and the halo anisotropy parameter $\beta$.

For the density slopes, we adopt the smooth halo density profiles measured by \citet{Ibata2014}, who masked prominent tidal substructures and provided power-law fits in different metallicity bins. This choice is consistent with our kinematic analysis, which is designed to isolate the halo after accounting for tidal substructures and foreground contamination.
To choose the appropriate metallicity bin, we use the metallicity of the dynamically hot halo component inferred from our GMM-decomposed DESI sample. The resulting mean halo metallicity is $\mathrm{[Fe/H]}=-0.56\pm0.72$, broadly consistent with SPLASH measurements of the halo, which give median $\mathrm{[Fe/H]}$ values from $-0.4$ to $-1.5$ \citep{Gilbert2014}. We therefore adopt the $-1.1<\mathrm{[Fe/H]}<0$ bin from \citet{Ibata2014}, for which $\alpha_{\rm 3D}=-3.66\pm0.21$ and $\alpha_{\rm 2D}=-2.66\pm0.21$. Given these slopes, we derive the normalization ratio following \citet{Zhang2024}, obtaining $\rho_0/\Sigma_0=0.59\,{\rm kpc}^{-1}$.

Finally, we fix the halo anisotropy parameter $\beta$ rather than fitting it as a free parameter. This choice is motivated by the well-known mass--anisotropy degeneracy, in which the inferred mass profile depends sensitively on the assumed anisotropy \citep{Merrifield1990,Wilkinson2002,Lokas2003}. Because there is currently no available direct constraint on $\beta$ for the M31 halo, we instead explore several representative constant values guided by measurements of the MW stellar halo.
Observational studies of MW halo stars generally find radially biased anisotropy. Analyses without explicit substructure removal typically give $\beta\sim0.3$--$0.4$ \citep{Shen2022,Medina2025a,Medina2025b}, while studies that remove substructure members using phase-space information infer more radial values, $\beta\sim0.5$--$0.8$ (\citealt{Bird2021}; Li et al. in prep.). Motivated by this range, we adopt $\beta=0.3$, 0.5, and 0.7 as representative cases.

With $\alpha_{\rm 3D}$, $\alpha_{\rm 2D}$, $\rho_0/\Sigma_0$, and $\beta$ specified, the observed $\sigma_{\rm LOS}(R_p)$ profile can be used in Eq.~\ref{eq:sigmaLOS_rp} to constrain the circular velocity profile in the halo region.

\section{The Mass distribution of M31}
\label{sec:mass distribution}

\subsection{The gravitational potential model}\label{subsec:potential model}

Using the newly derived M31 disk and halo circular velocity profile $V_{\rm c}(R)$, we construct a parameterized gravitational potential model to infer the mass distribution of M31. The model includes three components: a bulge, a disk, and a dark matter halo. The total gravitational potential is given by the sum of the individual component potentials.

The bulge is described by a Hernquist potential \citep{Hernquist1990}, while the disk is described by a Miyamoto--Nagai potential \citep{Miyamoto1975}:
\begin{equation}
\Phi_{\text{Bulge}} = -\frac{GM_{\rm b}}{r+q},
\label{eq:BulgePotential}
\end{equation}
\begin{equation}
\Phi_{\text{Disk}} = -\frac{GM_{\rm d}}{\sqrt{R^2 + \left(a_{\rm d} + \sqrt{z^2 + b_{\rm d}^2}\right)^2}},
\label{eq:DiskPotential}
\end{equation}
where $G$ is the gravitational constant, $r$ is the spherical radius, and $(R,z)$ are cylindrical coordinates. The parameters $M_{\rm b}$ and $q$ are the bulge mass and scale length, while $M_{\rm d}$, $a_{\rm d}$, and $b_{\rm d}$ are the disk mass, scale length, and scale height. The ratio $b_{\rm d}/a_{\rm d}$ controls the disk flattening, with smaller values corresponding to a thinner disk.
The corresponding circular velocity contributions are:
\begin{align}
V_{\rm c,Bulge}^2(r) &= r \frac{{\rm d} \Phi_{\rm Bulge}}{{\rm d} r} \,, \label{eq:Vc_bulge} \\
V_{\rm c,Disk}^2(R)  &= R \frac{{\rm d} \Phi_{\rm Disk}}{{\rm d} R}  \,. \label{eq:Vc_disk}
\end{align}

For the dark matter halo, we consider three density profiles: the Navarro--Frenk--White (NFW) profile \citep{Navarro1996}, the Dekel--Zhao (DZ) profile \citep{Freundlich2020}, and the Einasto profile \citep{Einasto1965,Montenegro2012}.
The NFW profile is widely used in galaxy dynamics and cosmological simulations. It is specified by two free parameters, the halo mass $M_{200}$ and concentration $c$, with potential
\begin{equation}
\Phi_{\text{Halo,NFW}} = -\frac{GM_{200} \ln{(1+r\,c/r_{200})}}{g(c)\,r},
\label{eq:NFWPotential}
\end{equation}
where $g(c)=\ln(1+c)-c/(1+c)$. Here $r_{200}$ is the radius within which the mean density is $200\rho_{\rm c}$, and $M_{200}$ is the enclosed mass within this radius. We define the critical density as
$\rho_{\rm c}=3H_0^2/(8\pi G)$ and adopt
$H_0=70\,{\rm km\,s^{-1}\,Mpc^{-1}}$ \citep{Komatsu2011}. The corresponding circular velocity contribution is
\begin{align}
V_{\rm c,Halo,NFW}^2(r) = r\frac{{\rm d}\Phi_{\rm Halo,NFW}}{{\rm d}r}.
\label{eq:Vc_halo_nfw}
\end{align}

As a more flexible alternative, we also consider the DZ profile, a special case of the Zhao double power-law family \citep{Zhao1996}. The DZ profile generalizes the NFW form by introducing an additional parameter that controls the inner density slope, allowing halos to range from cuspy to cored. 
It has been shown to provide better fits to dark matter density profiles in the hydrodynamical simulations than other profiles, with particularly good recovery of the corresponding circular velocity profiles, while retaining analytic expressions for the potential and velocity dispersion \citep{Freundlich2020}. Unlike the NFW profile, the DZ profile has a fixed outer asymptotic density slope of $\rho\propto r^{-3.5}$ \citep{Freundlich2020}, which is sufficiently steep for the enclosed mass to converge at large radii. These properties make the DZ profile well suited for mass modeling based on RC constraints.
The circular velocity profile of the DZ halo is \citep{Freundlich2020}
\begin{equation}
V_{\rm c,Halo,DZ}^2(r) = c_1\mu V_{200}^2\frac{x^{2-a}}{(1+x^{1/2})^{2(3-a)}},
\label{eq:dekel-zhao profile}
\end{equation}
where $x=r/r_{\rm c}$, $r_{\rm c}=r_{200}/c_1$, and
$\mu=c_1^{a-3}(1+c_1^{1/2})^{2(3-a)}$.
Here $V_{200}^2=GM_{200}/r_{200}$, with $M_{200}$ and $r_{200}$ defined as in the NFW model.
The DZ model has three free parameters: the halo mass $M_{200}$, the concentration-like parameter $c_1$, and the inner asymptotic slope $a$. The parameter $c_1$ is distinct from the standard NFW concentration. For comparison with NFW halo parameters, we also report two derived quantities: the inner logarithmic slope $s_1$ at $r_1=0.01r_{200}$, and the concentration $c=r_{200}/r_{-2}$, where $r_{-2}$ is the radius at which $\frac{d\ln\rho}{ d\ln r}=-2$. These are given by $s_1 = \frac{a+3.5 c_1^{1/2}(r_1/r_{200})^{1/2}}{1+c_1^{1/2}(r_1/r_{200})^{1/2}}$ and $c = c_1(\frac{1.5}{2-a})^2$, with $0<a<2$ required for physical validity.
For reference, an NFW profile with $c=15$ corresponds to $s_1\simeq1.25$ \citep{Freundlich2020}.

We also consider the Einasto profile \citep{Einasto1965,Montenegro2012}, which exhibits an exponential density decline that becomes progressively steeper with increasing radius, in contrast to the NFW and DZ profiles, whose outer density distributions asymptotically approach power laws with slopes of $-3$ and $-3.5$, respectively. We include the Einasto model to assess the dependence of the inferred M31 mass distribution on the adopted halo density profile.
The Einasto density profile is given by
\begin{equation}
\rho_{\rm Halo,Ein}(r)=\rho_{-2}
\exp\left[-\frac{2}{\alpha}
\left(\left(\frac{r}{r_{-2}}\right)^\alpha-1\right)\right],
\end{equation}
where $r_{-2}$ is the radius at which the logarithmic density slope equals $-2$, $\rho_{-2}$ is the density at this radius, and $\alpha$ is the shape parameter. 
The enclosed mass is
\begin{equation}
\begin{aligned}
M_{\rm Halo,Ein}(<r)={}&
4\pi \rho_{-2} r_{-2}^{3}\frac{e^{2/\alpha}}{\alpha}\left(\frac{\alpha}{2}\right)^{3/\alpha} \\
&\times\gamma\left[\frac{3}{\alpha},\frac{2}{\alpha}\left(\frac{r}{r_{-2}}\right)^\alpha\right].
\end{aligned}
\end{equation}
where $\gamma(a,x)$ denotes the lower incomplete gamma function
\citep{Montenegro2012}.
For consistency with the NFW and DZ models, we reparameterize the Einasto model in terms of $M_{200}$, $c$, and $\alpha$. The scale radius is related to the concentration through $r_{-2}=r_{200}/c$, while $\rho_{-2}$ is obtained by imposing $M_{\rm Halo,Ein}(<r_{200})=M_{200}$. This gives
\begin{equation}
\rho_{-2} = \frac{ M_{200}\,\alpha e^{-2/\alpha}(2/\alpha)^{3/\alpha}}{4\pi r_{-2}^{3}\gamma\left[
3/\alpha,(2/\alpha)c^\alpha\right]}.
\end{equation}
Substituting these relations into the cumulative mass profile yields
\begin{equation}
M_{\rm Halo,Ein}(<r) = M_{200}\frac{ \gamma\left[3/\alpha,(2/\alpha) \left(c r/r_{200}\right)^\alpha\right]}{\gamma\left[3/\alpha,(2/\alpha)c^\alpha\right]}.
\end{equation}
The corresponding circular velocity is
\begin{equation}
V_{\rm c,Halo,Ein}^2(r)=\frac{G M_{\rm Halo,Ein}(<r)}{r}.
\end{equation}

Combining the contributions from the bulge, disk, and dark matter halo, the total circular velocity profile of M31 is:
\begin{equation}\label{eq:Vc_total_nfw}
V_{\rm c,model}^2 = V_{\rm c,Bulge}^2 + V_{\rm c,Disk}^2 + V_{\rm c,Halo}^2\,.
\end{equation}
For the baryonic components, we vary only the disk mass $M_{\rm d}$, while the remaining structural parameters are fixed to the values listed in Table~\ref{tab:Potential_model_para}. The bulge parameters are adopted from \citet{Courteau2011} and \citet{Tamm2012}, while the disk scale length and scale height are taken from \citet{Dalcanton2023}, consistent with earlier studies of M31's thick disk \citep{Dalcanton2015,Dorman2015}. We impose an upper limit of $M_{\rm d}<10^{11}\,M_{\odot}$, motivated by previous dynamical estimates and the relatively well-constrained disk mass-to-light ratio of M31 \citep[e.g.,][]{Chemin2009,Corbelli2010}. This prior prevents the disk contribution from becoming unrealistically large and reduces the disk--halo degeneracy in the RC fit. For the dark matter halo, the free parameters are $(M_{200},c)$ for the NFW model, $(M_{200},c_1,a)$ for the DZ model, and $(M_{200},c,\alpha)$ for the Einasto model.

\begin{table}
	\centering
	\caption{Parameters of the gravitational potential model for M31. Uniform priors are adopted for all free parameters.}
	\label{tab:Potential_model_para}
    \resizebox{0.49\textwidth}{!}{ 
	\begin{tabular}{llcl} 
	\hline
	Component & Parameter & Prior & Note \\
	\hline
	\multirow{2}{*}{Bulge}  
        & $M_{\rm b}$   & $3.1 \times 10^{10}\,M_{\odot}$ & Fixed \\
	    & $q$           & $0.7$ kpc & Fixed \\
	\hline
	\multirow{3}{*}{Disk}    
        & $M_{\rm d}$   & $10 < \log_{10}(M_{\rm d}/M_{\odot}) < 11$ & Free \\
        & $a_{\rm d}$   & $5.5$ kpc & Fixed \\
        & $b_{\rm d}$   & $0.77$ kpc & Fixed \\
	\hline
	\multirow{2}{*}{NFW Halo}  
        & $M_{200}$     & $0.01 < M_{200}/10^{12}\,M_{\odot} < 100$ & Free \\
	    & $c$           & $0.1 < \log_{10} c < 2.0$ & Free \\
    \cline{2-4}
	\multirow{3}{*}{DZ Halo}    
        & $M_{200}$     & $0.01 < M_{200}/10^{12}\,M_{\odot} < 100$ & Free \\
        & $c_1$         & $0.1 < \log_{10} c_1 < 2.0$ & Free \\
        & $a$           & $0 < a < 2$ & Free \\
    \cline{2-4}    
    \multirow{3}{*}{Einasto Halo}  
        & $M_{200}$     & $0.01 < M_{200}/10^{12}\,M_{\odot} < 100$ & Free \\
	    & $c$           & $0.1 < \log_{10} c < 2.0$ & Free \\
	    & $\alpha$           & $0.01 < \alpha < 2.0$ & Free \\
	\hline
	\end{tabular}
    }
\end{table}

\begin{table*}
    \centering
    \caption{Best-fit parameters for the NFW, Dekel--Zhao (DZ), and Einasto halo profiles under different assumed values of the halo velocity anisotropy $\beta$. For all models, we report the disk mass $M_{\rm d}$, the dark matter halo mass $M_{200}$, and the concentration $c=r_{200}/r_{-2}$. For the DZ model, $c$ and the inner logarithmic slope $s_1$ at $r_1=0.01\,r_{200}$ are derived from the fitted parameters $c_1$ and $a$, whereas $\alpha$ denotes the fitted shape parameter of the Einasto profile. The reported parameter values are the medians of the posterior distributions, with uncertainties corresponding to the 16th and 84th percentiles. We also report the chi-squared values per data point for the disk and halo data, $\chi^2_{\rm disk}/N_{\rm disk}$ and $\chi^2_{\rm halo}/N_{\rm halo}$, respectively, together with the reduced chi-square $\chi^2_\nu$ of the joint fit. The radius $r_{200}$ is calculated from $M_{200}=(4\pi/3)\,200\,\rho_{\rm c}\,r_{200}^{3}$. The quantity $M_{\rm tot}(<r_{200})$ denotes the total mass of the baryonic and dark matter components enclosed within $r_{200}$.}
    \label{tab:mass results}

    \small
    \setlength{\tabcolsep}{3pt}
    \begin{tabular*}{\textwidth}{@{\extracolsep{\fill}}lccccccccccc}
        \hline
        Halo profile & $\beta$ & \multicolumn{5}{c}{Mass-model results} & \multicolumn{3}{c}{Goodness of fit} & \multicolumn{2}{c}{Calculated results} \\
        \cline{3-7} \cline{8-10} \cline{11-12}
        & & $M_{\rm d}$ & $M_{200}$ & $\log c$ & $s_1$ & $\alpha$ & $\chi^2_{\rm disk}\over N_{\rm disk}$ & $\chi^2_{\rm halo} \over N_{\rm halo}$ & $\chi^2_\nu$ & $r_{200}$ & $M_{\rm tot}(<r_{200})$ \\
        & & $(10^{10}\,M_{\odot})$ & $(10^{12}\,M_{\odot})$ & & & & & & & (kpc) & $(10^{12}\,M_{\odot})$ \\
        \hline

        \multirow{3}{*}{NFW}
        & 0.7 & $6.92_{-2.94}^{+1.99}$ & $1.37_{-0.14}^{+0.16}$ & $1.19_{-0.08}^{+0.09}$ & -- & -- & 1.70 & 0.90 & 1.58 & $229.10_{-8.09}^{+8.60}$ & $1.47_{-0.14}^{+0.17}$ \\
        & 0.5 & $7.59_{-2.69}^{+1.53}$ & $1.15_{-0.12}^{+0.13}$ & $1.21_{-0.07}^{+0.08}$ & -- & -- & 2.04 & 1.61 & 2.11 & $216.12_{-7.79}^{+7.85}$ & $1.25_{-0.11}^{+0.13}$ \\
        & 0.3 & $8.51_{-2.05}^{+1.04}$ & $1.00_{-0.11}^{+0.12}$ & $1.22_{-0.06}^{+0.07}$ & -- & -- & 2.37 & 2.44 & 2.68 & $206.28_{-7.86}^{+7.94}$ & $1.12_{-0.10}^{+0.11}$ \\
        \hline

        \multirow{3}{*}{DZ}
        & 0.7 & $7.76_{-2.51}^{+1.57}$ & $1.52_{-0.17}^{+0.21}$ & $1.15_{-0.07}^{+0.09}$ & $1.23_{-0.07}^{+0.08}$ & -- & 1.66 & 0.91 & 1.62 & $237.18_{-9.19}^{+10.45}$ & $1.63_{-0.17}^{+0.21}$ \\
        & 0.5 & $8.32_{-2.15}^{+1.23}$ & $1.26_{-0.14}^{+0.16}$ & $1.18_{-0.06}^{+0.08}$ & $1.26_{-0.07}^{+0.08}$ & -- & 1.98 & 1.71 & 2.20 & $222.80_{-8.58}^{+9.06}$ & $1.37_{-0.14}^{+0.17}$ \\
        & 0.3 & $8.91_{-1.67}^{+0.86}$ & $1.10_{-0.12}^{+0.14}$ & $1.20_{-0.05}^{+0.06}$ & $1.27_{-0.06}^{+0.07}$ & -- & 2.36 & 2.52 & 2.81 & $212.94_{-8.04}^{+8.68}$ & $1.21_{-0.12}^{+0.14}$ \\
        \hline

        \multirow{3}{*}{Einasto}
        & 0.7 & $7.50_{-3.05}^{+1.81}$ & $0.76_{-0.15}^{+0.20}$ & $1.11_{-0.06}^{+0.07}$ & -- & $0.39_{-0.07}^{+0.09}$ & 1.35 & 1.23 & 1.54 & $188.25_{-13.30}^{+15.25}$ & $0.86_{-0.16}^{+0.21}$ \\
        & 0.5 & $6.75_{-3.58}^{+2.33}$ & $0.48_{-0.10}^{+0.14}$ & $1.14_{-0.06}^{+0.07}$ & -- & $0.51_{-0.11}^{+0.15}$ & 1.30 & 2.28 & 1.98 & $164.81_{-12.82}^{+14.79}$ & $0.58_{-0.10}^{+0.14}$ \\
        & 0.3 & $4.79_{-2.94}^{+3.44}$ & $0.34_{-0.06}^{+0.08}$ & $1.17_{-0.07}^{+0.05}$ & -- & $0.72_{-0.18}^{+0.25}$ & 0.93 & 3.83 & 2.42 & $143.97_{-9.02}^{+10.51}$ & $0.41_{-0.06}^{+0.09}$ \\
        \hline
    \end{tabular*}
\end{table*}

\begin{figure*}
\centering
\includegraphics[width=0.32\textwidth]{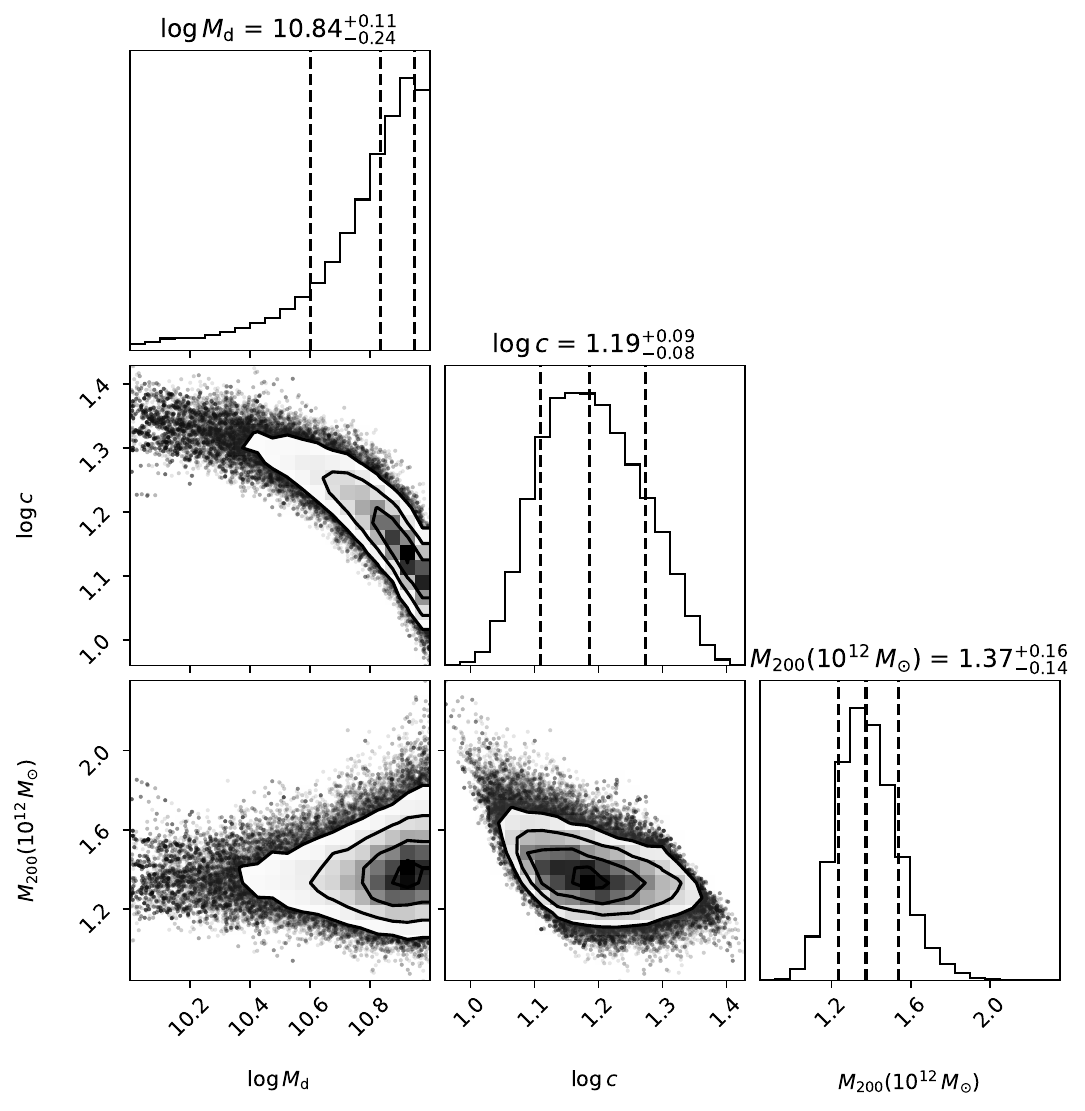}
\hfill
\includegraphics[width=0.32\textwidth]{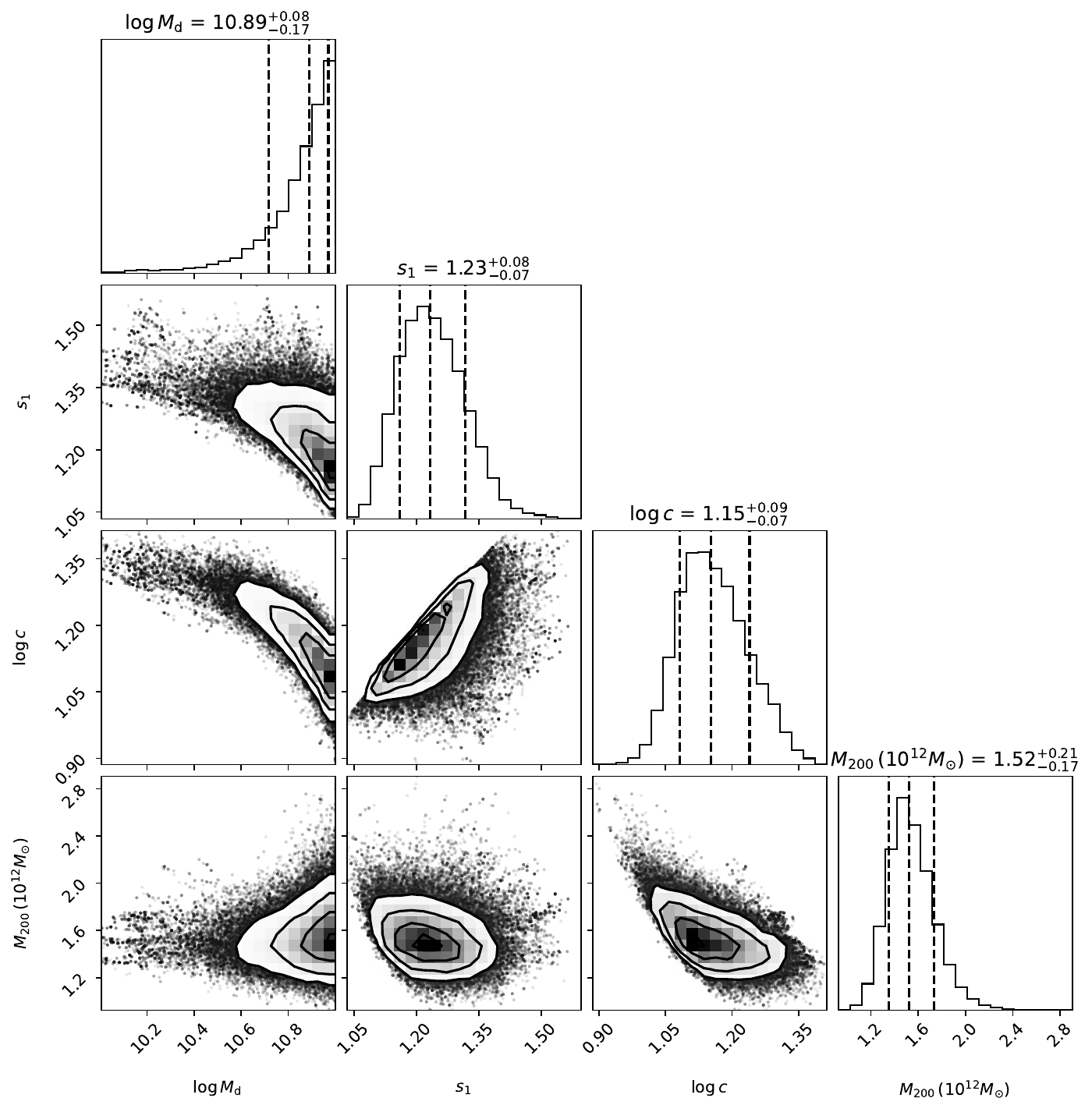}
\hfill
\includegraphics[width=0.32\textwidth]{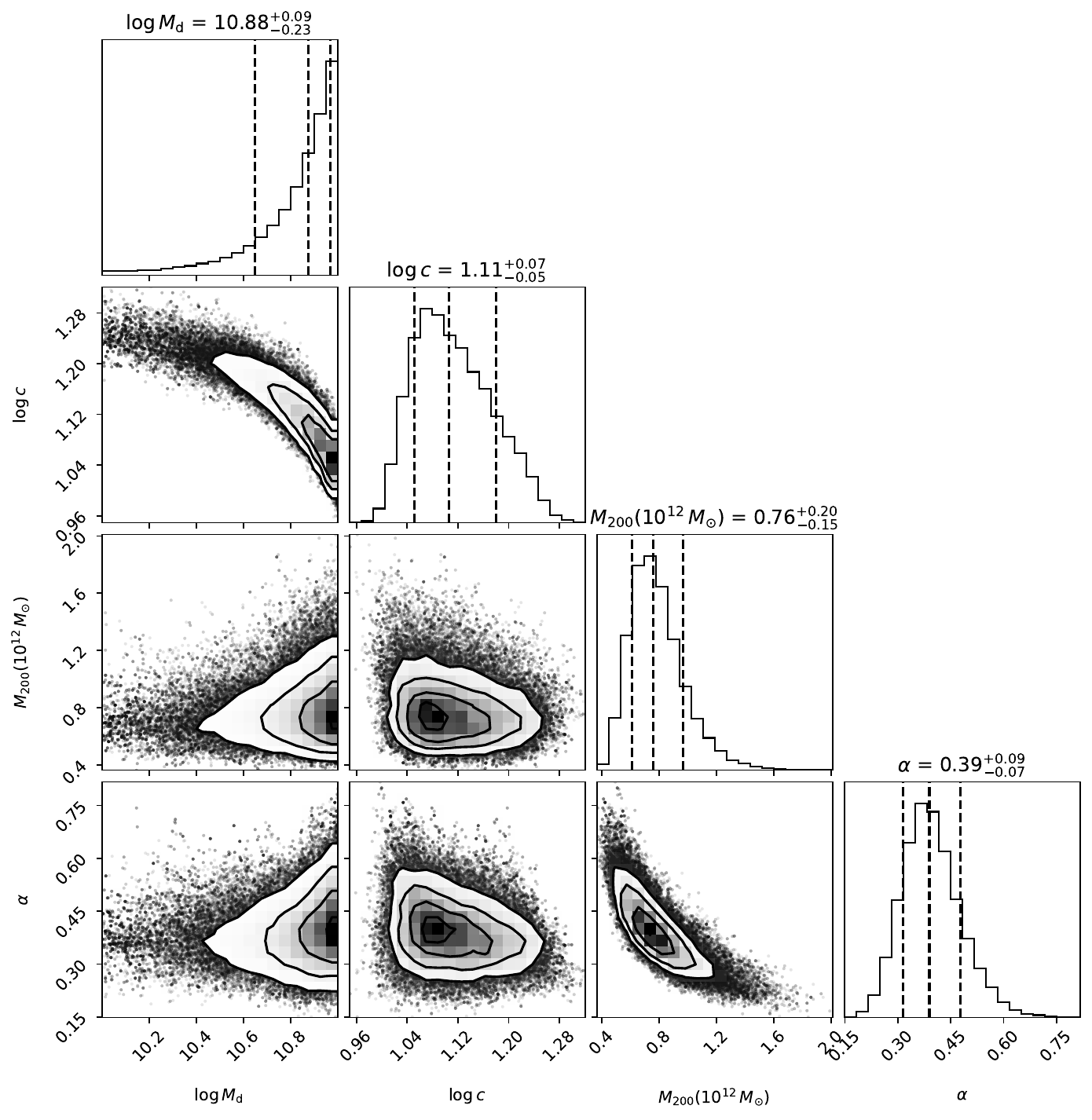}
\caption{Marginalized probability distribution functions (PDFs) of the mass-model parameters derived from the MCMC analysis for $\beta=0.7$. The left, middle, and right panels show the results for the NFW, DZ, and Einasto halo models, respectively. The corner plots show the two-dimensional marginalized PDFs, with contours indicating the $1\sigma$, $2\sigma$, and $3\sigma$ confidence regions. The histograms along the diagonal show the one-dimensional marginalized PDFs of each parameter, with dashed lines marking the 16th, 50th, and 84th percentiles. The median values and corresponding uncertainties are listed above each one-dimensional distribution.}
\label{fig:posterior_pdf}
\end{figure*}

With the parameterized potential models specified above, we jointly fit two sets of observables: the disk circular velocity profile, $V_{\rm c}(R)$, and the halo LOS velocity dispersion profile, $\sigma_{\rm LOS}^{\rm Halo}(R_p)$. The disk RC directly constrains the model circular velocity, while the halo dispersion profile constrains the same potential through Eq.~\ref{eq:sigmaLOS_rp}, using the parameters determined in Sections~\ref{subsec:halo vel disp} and \ref{subsec:halo parameter}.
We sample the posterior distributions of the free parameters using the MCMC sampler \texttt{emcee} \citep{emcee}. The fitted parameters are $M_{\rm d}$, $M_{200}$, and $c$ for the
NFW halo, $M_{\rm d}$, $M_{200}$, $c_1$, and $a$ for the DZ halo, and $M_{\rm d}$, $M_{200}$, $c$, and $\alpha$ for the Einasto halo. The likelihood function is
\begin{align}
\label{eq:likelihood}
\ln \mathcal{L}(\theta)
= &-\frac{1}{2} \Biggl[
\sum_{k=1}^{n}
\frac{
\left[
V_{{\rm c},k}^{\rm obs}
-
V_{\rm c}^{\rm model}(R_k|\theta)
\right]^2
}{
\sigma_{V_{{\rm c},k}}^2
}
\notag\\
&+
\sum_{j=1}^{m}
\frac{
\left[
\sigma_{{\rm LOS},j}^{\rm Halo,obs}
-
\sigma_{\rm LOS}^{\rm Halo}
\left(R_{p,j}; V_{\rm c}^{\rm model}(\theta), p\right)
\right]^2
}{
\sigma_{\sigma_{{\rm LOS},j}^{\rm Halo}}^2
}
\Biggr].
\end{align}
Here, $\theta$ denotes the fitted potential model parameters. $V_{\rm c}^{\rm model}(R_k|\theta)$ is the model-predicted circular velocity at radius $R_k$ from the adopted potential model, while $\sigma_{\rm LOS}^{\rm Halo}(R_{p,j}; V_{\rm c}^{\rm model}(\theta), p)$ denotes the projected halo LOS velocity dispersion predicted by inserting the circular velocity model into Eq.~\ref{eq:sigmaLOS_rp}. 
Here $p$ denotes the fixed halo parameters entering Eq.~\ref{eq:sigmaLOS_rp} determined in Sections~\ref{subsec:halo vel disp} and~\ref{subsec:halo parameter}, including the halo dispersion slope, the 3D and 2D density slopes, the normalization ratio, and the anisotropy parameter.
The observed quantities, $V_{{\rm c},k}^{\rm obs}$ and $\sigma_{{\rm LOS},j}^{\rm Halo,obs}$, are taken from Tables~\ref{tab:RC_disk} and~\ref{tab:halo_dispersion}, respectively. Their uncertainties are denoted by $\sigma_{V_{{\rm c},k}}$ and $\sigma_{\sigma_{{\rm LOS},j}^{\rm Halo}}$, and $n$ and $m$ are the numbers of disk RC and halo dispersion data points, respectively.
We exclude disk data at $R<4\,{\rm kpc}$ from the fit because the observed $V_{\rm c}$ in this region may be affected by bar-induced perturbations and other non-axisymmetric motions, which can drive deviations from circular orbits and make the inner RC an unreliable tracer of the underlying gravitational potential \citep[e.g.,][]{Chemin2009,Blana2018,Liu2025}.

\subsection{Results of Mass Modeling} 
\label{subsec:mass_results}

We infer the mass distribution of M31 using the NFW, DZ, and Einasto halo models for the three assumed anisotropy values, $\beta=0.3$, 0.5, and 0.7. The fitting results and goodness-of-fit statistics are summarized in Table~\ref{tab:mass results}, including the chi-squared values per data point for the disk and halo and the reduced chi-squared value of the joint fit. For comparison with the NFW model, we convert the DZ parameters $c_1$ and $a$ into the equivalent concentration $c$ and inner logarithmic slope $s_1$, following the definitions in Section~\ref{subsec:potential model}. Among the three adopted anisotropy values, the $\beta=0.7$ cases yield the smallest reduced chi-squared values for all three halo profiles. The marginalized posterior probability distributions for the three halo models in the $\beta=0.7$ cases are shown in Fig.~\ref{fig:posterior_pdf}.

\begin{figure*}
\centering
    \includegraphics[width=\textwidth]{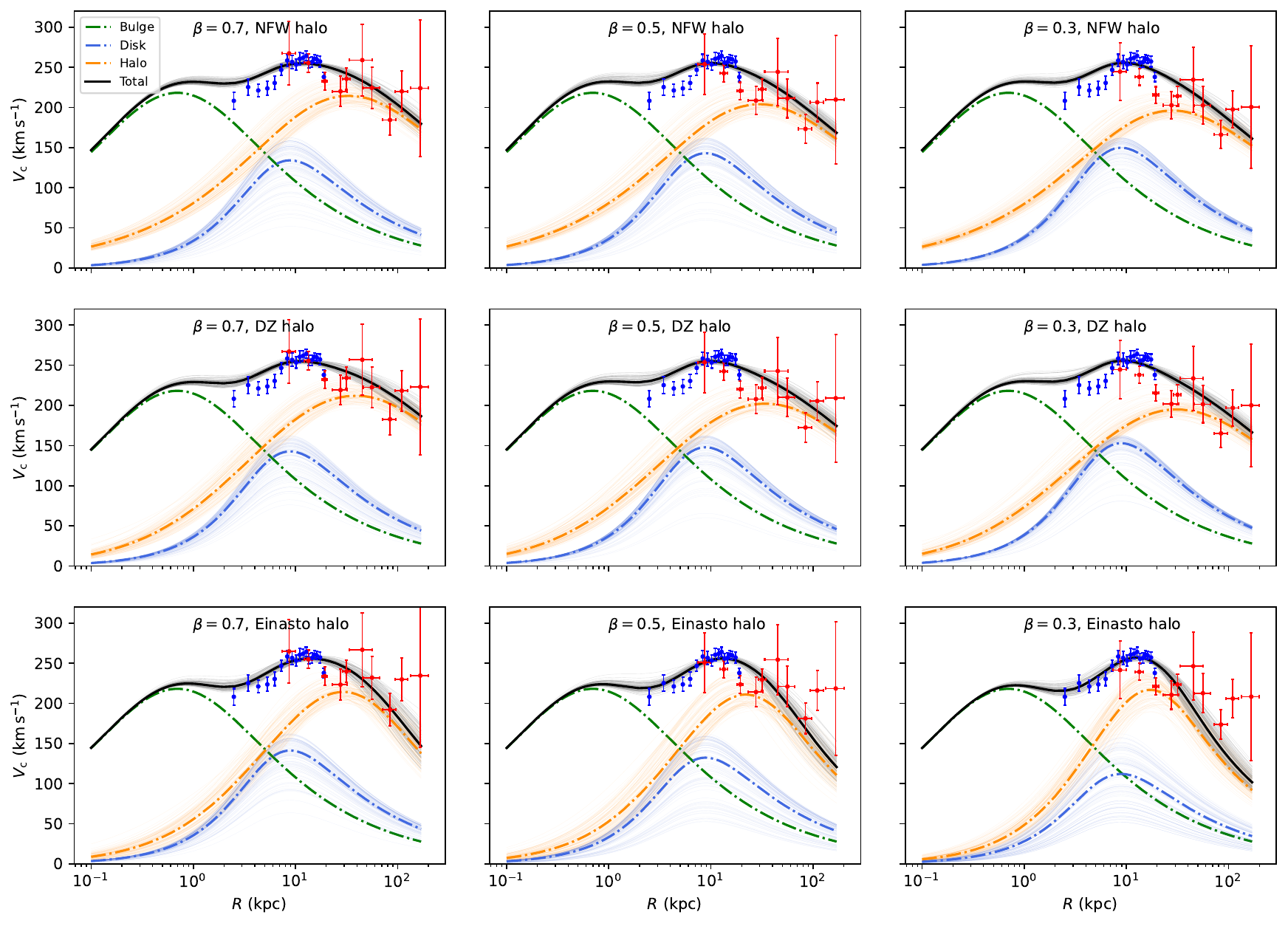}
\caption{Rotation curves of M31 for different $\beta$ values and halo models. The adopted $\beta$ value and halo model are noted in the top middle corner of each panel. The total circular velocity is shown as the black solid curve, while the contributions from the bulge, disk, and dark matter halo are plotted as green, blue, and orange dash–dotted curves, respectively. Semi-transparent curves show 100 posterior draws from the model parameter distributions, illustrating the associated uncertainties. Observed data points in the disk and halo regions, along with their uncertainties, are shown as blue and red symbols, respectively.}
\label{fig:RC_combine}
\end{figure*}

The RCs derived from the best-fit models for different $\beta$ values and halo profiles are shown in Fig.~\ref{fig:RC_combine}, with the observed disk circular velocities overplotted for comparison. The halo data are measured as projected dispersions, $\sigma_{\rm LOS}^{\rm Halo}(R_p)$, rather than as direct circular velocities. For visualization only, we convert each of the ten halo dispersion measurements into an effective circular velocity through Eq.~\ref{eq:sigmaLOS_rp} by setting $R=R_p$ and using the best-fit model parameters. In the actual fit, the halo dispersion profile is compared directly with the model prediction in projected space.

\begin{figure*}
  \centering
\includegraphics[width=\textwidth]{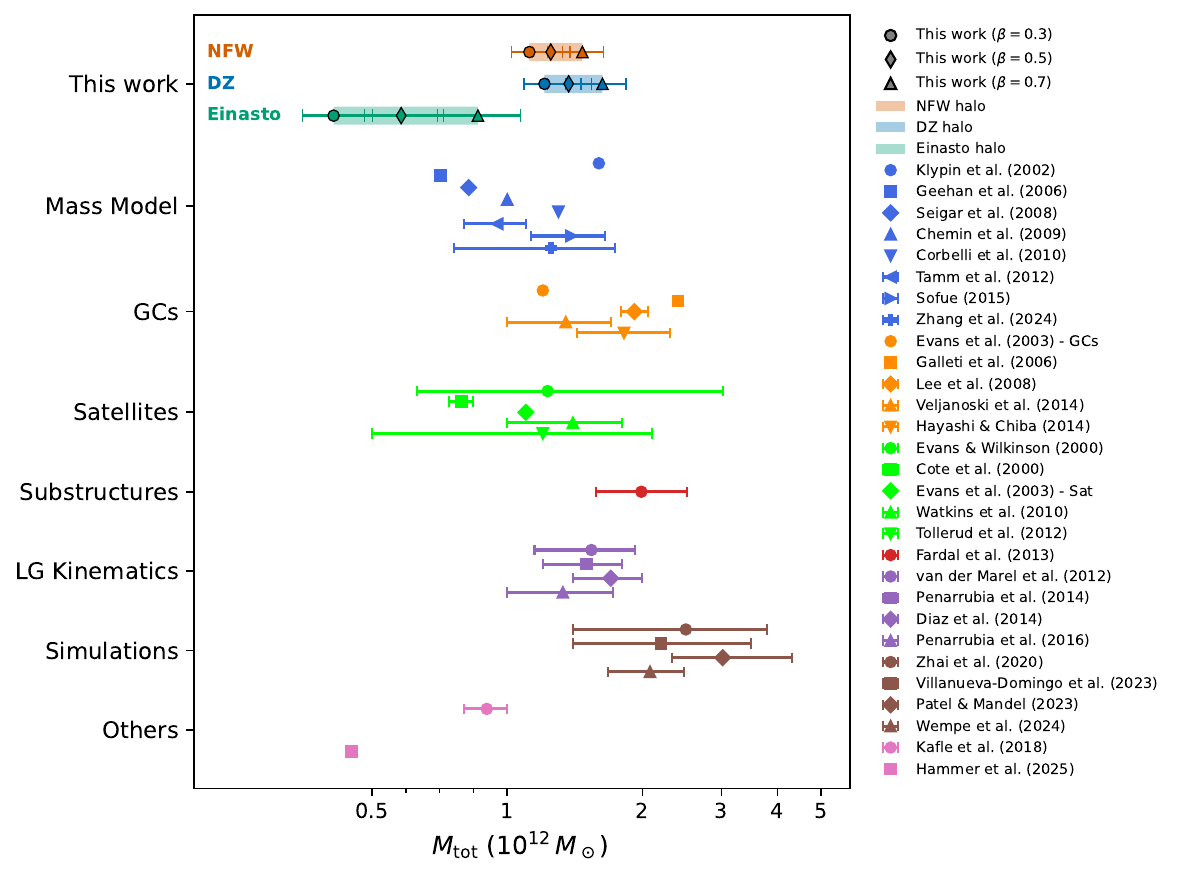}
    \caption{Comparison of M31 total-mass measurements published in the 21st century. We include measurements that either report the total mass or enclosed mass at radius beyond 200 kpc. The results from this work are shown at the top. Colors distinguish the adopted dark matter halo profiles, with orange representing NFW, blue representing the DZ profile, and green representing Einasto. Marker shapes distinguish the assumed halo velocity anisotropy values, $\beta=0.3$, $0.5$, and $0.7$. The horizontal shaded bands span the range of median total masses obtained for $0.3\leq\beta\leq0.7$ for each halo profile, illustrating the systematic variation associated with the assumed anisotropy. The error bars on the individual measurements indicate their corresponding formal posterior uncertainties. Literature measurements are grouped according to method and shown in different colors, with blue representing RC-based mass modeling, orange representing GC kinematics, green representing satellite kinematics, red representing substructure-based estimates, purple representing Local Group kinematics, brown representing cosmological simulations, and pink representing other methods. Within each category, individual literature measurements are distinguished by different marker shapes.}
  \label{fig:Mass_compare_total}
\end{figure*}

\begin{figure}
  \centering
    \centering
    \includegraphics[width=\linewidth]{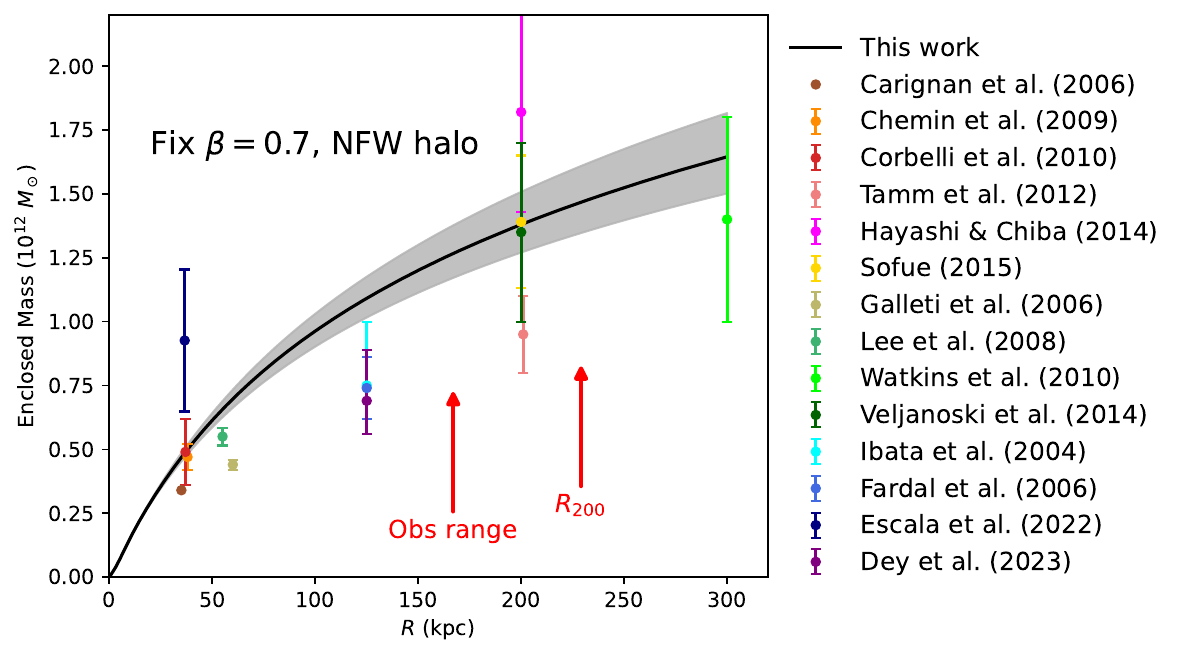}
  \caption{The cumulative mass profiles $M(<R)$ derived from the best-fit model assuming $\beta = 0.7$ for the NFW halo. Black curve shows the best-fitting profile, with the gray shaded region indicating the associated $1\sigma$ uncertainty. For comparison, enclosed mass estimates reported in other studies are overplotted at their respective radii.}
  \label{fig:Mass_compare_enclosed}
\end{figure}

At fixed $\beta$, the NFW and DZ models provide comparably good fits to the data, and their inferred parameters are generally consistent within $1\sigma$, although the DZ model tends to favor slightly larger disk and halo masses. Across these two halo profiles and the three adopted anisotropy values, the inferred total mass within $r_{200}$ ranges from $M_{\rm tot}(<r_{200})=1.12_{-0.10}^{+0.11}\times10^{12}\,M_{\odot}$ for the NFW halo with $\beta=0.3$ to $1.63_{-0.17}^{+0.21}\times10^{12}\,M_{\odot}$ for the DZ halo with
$\beta=0.7$. As shown in Fig.~\ref{fig:Mass_compare_total}, this range is broadly consistent with most previous observational estimates obtained using different methods \citep[e.g.,][]{Chemin2009,Corbelli2010,Watkins2010,vdm2012a,
Tollerud2012,Penarrubia2014,Penarrubia2016,Sofue2015,Zhang2024}.
For both profiles, the inferred RC reaches a maximum circular velocity of approximately $255\,{\rm km\,s^{-1}}$ at $R\sim10\,{\rm kpc}$ and declines gradually toward larger radii. At the outer edge of our observational range, $R=167\,{\rm kpc}$, the best-fit circular velocities are approximately $185$, $170$, and $165\,{\rm km\,s^{-1}}$ for $\beta=0.7$, $0.5$, and $0.3$, respectively. For illustration, we present a representative case using the NFW model with $\beta=0.7$, listing its effective halo circular velocities in Table~\ref{tab:halo_vc}. The corresponding cumulative mass profile is shown in Fig.~\ref{fig:Mass_compare_enclosed}, together with previous enclosed mass estimates at specific radii.

\begin{table}
    \centering
	\caption{Transformed observational circular velocity in M31's halo region for the best-fit NFW model with $\beta = 0.7$. The values are inferred from the observed $\sigma_{\rm LOS}^{\rm Halo}(R_p)$ under the approximation $R = R_p$. This approximation is not used in the fitting itself, where the observed 2D $\sigma_{\rm LOS}^{\rm Halo}(R_p)$ is directly used to constrain the mass distribution.}
	\label{tab:halo_vc}
	\begin{tabular}{cccc}
	\hline
	$R$ (kpc) & $\sigma_{R}$ (kpc)  & $V_{\rm c}\,\rm(km\,s^{-1})$ & $\sigma_{V_{\rm c}}\,\rm(km\,s^{-1})$ \\ 
	\hline
	8.66 & 1.26 & 267.29 & 39.58 \\
    13.38 & 1.16 & 255.38 & 11.38 \\
	19.37 & 1.21 & 232.73 & 11.13 \\
    27.62 & 4.59 & 219.99 & 18.66 \\
	31.73 & 2.62 & 235.58 & 13.54 \\
    45.13 & 11.48 & 258.91 & 44.53 \\
	56.10 & 10.25 & 224.29 & 25.77 \\
	84.30 & 12.50 & 184.42 & 19.31 \\
	109.97 & 15.75 & 219.98 & 25.49 \\
	166.80 & 33.50 & 223.97 & 85.30 \\
	\hline
	\end{tabular}
\end{table}

The Einasto model, however, yields a substantially different and much lower mass scale than the NFW and DZ models. As shown in Fig.~\ref{fig:Mass_compare_total}, across the three adopted anisotropy values, the inferred total mass within $r_{200}$ ranges from $M_{\rm tot}(<r_{200})=0.41_{-0.06}^{+0.09}\times10^{12}\,M_{\odot}$ for $\beta=0.3$ to $0.86_{-0.16}^{+0.21}\times10^{12}\,M_{\odot}$ for $\beta=0.7$. This marked difference reflects the dependence of the inferred virial mass on the adopted halo density profile. In contrast to the NFW and DZ profiles, whose outer density slopes asymptotically approach fixed power laws, the logarithmic slope of the Einasto profile continues to steepen with increasing radius. Similar profile dependence has been reported in studies of the MW RC \citep{Salas2019,Jiao2021,Ou2024}. Additionally, \citet{Ou2025} showed that, even when the observational data no longer require a rapid decline in the outer MW RC, the Einasto model still favors a very low $M_{200}$, whereas the NFW model allows a substantially larger virial mass with a comparably acceptable fit.

A similar pattern is evident in our M31 analysis. While all three halo profiles provide acceptable fits to the combined disk and halo data, the Einasto model provides a noticeably poorer fit to the halo kinematic measurements than the NFW and DZ models at fixed $\beta$, as indicated by its larger $\chi^2_{\rm halo}\over N_{\rm halo}$ in Table~\ref{tab:mass results}. The corresponding differences in the outer RCs are illustrated in Fig.~\ref{fig:RC_combine}. The halo measurements provide the most direct constraints in our analysis on the mass distribution at large radii. Therefore, within the equilibrium-based framework adopted here, the poorer agreement of the Einasto model with these data suggests that the halo kinematics do not favor the rapid decline of the outer halo density associated with its low-$M_{200}$ solution.
Furthermore, the low virial mass inferred from the Einasto model is in tension with independent mass estimates based on M31 satellite galaxies, which generally favor a more massive halo and are more consistent with the NFW and DZ results \citep[e.g.,][]{Evans&Wilkinson2000,Cote2000,Evans2003,Watkins2010,Tollerud2012}. Taken together, the current data do not provide compelling evidence for preferring the low-mass Einasto solution.

\begin{table*}
\centering
\caption{Results of the outer-radius test for the NFW model with $\beta=0.7$. For each maximum projected radius $R_{p,\rm max}$ of the halo sample, we list the corresponding power-law index $\gamma$ of the halo LOS dispersion profile, inferred halo concentration, halo mass $M_{200}$, and total enclosed masses $M_{\rm tot}(<r)$ at three representative radii. The $R_{p,\rm max}=166.80$\,kpc case corresponds to the fiducial analysis.}
\label{tab:rmax_test}
\small
\setlength{\tabcolsep}{3pt}
\renewcommand{\arraystretch}{1.15}
\begin{tabular*}{\textwidth}{@{\extracolsep{\fill}}lcccccc@{}}
\hline
$R_{p,\rm max}$ & $\gamma$ & $\log c$ & $M_{200}$ & \shortstack{$M_{\rm tot}$\\$(<56.10\,{\rm kpc})$} & \shortstack{$M_{\rm tot}$\\$(<84.30\,{\rm kpc})$} & \shortstack{$M_{\rm tot}$\\$(<109.97\,{\rm kpc})$} \\
(kpc) & & & ($10^{12}\,M_\odot$) & ($10^{12}\,M_\odot$) & ($10^{12}\,M_\odot$) & ($10^{12}\,M_\odot$) \\
\hline
56.10 & $-0.11\pm0.06$ & $1.18_{-0.08}^{+0.10}$ & $1.40_{-0.16}^{+0.20}$ & $0.67_{-0.04}^{+0.04}$ & -- & -- \\
84.30 & $-0.15\pm0.05$ & $1.20_{-0.08}^{+0.09}$ & $1.37_{-0.14}^{+0.16}$ & $0.66_{-0.03}^{+0.03}$ & $0.87_{-0.05}^{+0.06}$ & -- \\
109.97 & $-0.13\pm0.04$ & $1.19_{-0.08}^{+0.09}$ & $1.38_{-0.14}^{+0.17}$ & $0.67_{-0.03}^{+0.03}$ & $0.87_{-0.05}^{+0.06}$ & $1.02_{-0.07}^{+0.08}$ \\
166.80 (fiducial) & $-0.13\pm0.04$ & $1.19_{-0.08}^{+0.09}$ & $1.37_{-0.14}^{+0.16}$ & $0.67_{-0.03}^{+0.03}$ & $0.87_{-0.05}^{+0.06}$ & $1.02_{-0.07}^{+0.07}$ \\
\hline
\end{tabular*}
\end{table*}

As discussed in Section~\ref{subsec:halo vel disp}, the outermost halo measurements may be more susceptible to departures from dynamical equilibrium. To assess whether these measurements significantly affect the inferred M31 mass, we repeat the analysis after progressively restricting the maximum projected radius of the halo sample. We adopt the NFW model with $\beta=0.7$ as a representative case and consider $R_{p,\rm max}=56.10$, 84.30, 109.97, and 166.80\,kpc. For each radial cut, we first truncate the halo sample at the corresponding $R_{p,\rm max}$ and refit its LOS velocity dispersion profile using Eq.~\ref{eq:halo_los_dispersion}, obtaining the corresponding power-law index $\gamma$. We then use the fitted $\gamma$, together with the truncated halo dispersion measurements and the same disk RC data, to jointly constrain the mass distribution through Eq.~\ref{eq:likelihood}, following the same fitting procedure as in the fiducial analysis. The resulting $\gamma$, halo concentration, $M_{200}$, and enclosed masses are summarized in Table~\ref{tab:rmax_test}. The fitted $\gamma$, $M_{200}$, and enclosed masses remain mutually consistent within their uncertainties across the different radial cuts without any systematic variation. This indicates that the inferred mass distribution is insensitive to the inclusion of the outermost halo measurements.

\section{Summary}
\label{sec:Summary}

Using more than 8,500 objects with LOS velocity measurements, we construct the RC of M31 from 2 to 167\,kpc. In the disk region, we select nearly 2,000 robust disk tracers and derive the circular velocity profile using a Jeans-based kinematic model. The resulting RC is in good agreement with previous H~\textsc{I} measurements.
In the halo region, we use DESI RGB stars and carefully selected GCs as kinematic tracers. We apply a Gaussian mixture model to infer the LOS velocity dispersion of the dynamically hot stellar halo, mitigating contamination from Galactic foreground stars and M31 tidal substructures without overly rejecting genuine halo members. The resulting dispersion profile is in good agreement with that derived by \citet{Gilbert2018} from the SPLASH survey.
We then combine the velocity dispersion measurements from DESI stars, SPLASH stars, and GCs after accounting for identifiable substructures to construct a global halo LOS velocity dispersion profile as a function of projected radius. This profile shows a mild radial decline, with a logarithmic slope of $-0.13\pm0.04$. Using this dispersion profile, we apply a deprojection method to constrain the corresponding halo circular velocity profile.

Combining the disk and halo kinematic constraints, we construct a gravitational potential model consisting of a bulge, a disk, and a dark matter halo. To assess the impact of the mass--anisotropy degeneracy, we fix the halo velocity anisotropy parameter to three representative values, $\beta=0.3$, 0.5, and 0.7. For the dark matter halo, we consider three parameterizations: the NFW, DZ, and Einasto profiles. The NFW and DZ models yield mutually consistent mass estimates, with the inferred total mass within $r_{200}$ ranging from $1.12^{+0.11}_{-0.10}\times10^{12}\,M_\odot$ for the NFW halo with $\beta=0.3$ to $1.63^{+0.21}_{-0.17}\times10^{12}\,M_\odot$ for the DZ halo with $\beta=0.7$.

The Einasto profile, in contrast, yields substantially lower masses, ranging from $0.41^{+0.09}_{-0.06}\times10^{12}\,M_\odot$ for $\beta=0.3$ to $0.86^{+0.21}_{-0.16}\times10^{12}\,M_\odot$ for $\beta=0.7$, demonstrating that the inferred virial mass remains sensitive to the adopted functional form of the dark matter halo. Although the Einasto model provides an acceptable fit to the combined data, it gives a poorer description of the halo kinematic measurements than the NFW and DZ models at fixed $\beta$, and its low-mass solutions are in tension with independent mass estimates from M31 satellites. The current data therefore do not provide compelling evidence for preferring the low-mass Einasto solution.

Overall, after carefully accounting for identifiable tidal substructures and Galactic foreground contamination, our analysis uses the residual dynamically hot halo to provide a comprehensive measurement of M31's mass distribution under the equilibrium assumption, while quantifying systematic uncertainties associated with halo velocity anisotropy and dark matter profile choice.

In the future, ongoing and upcoming surveys such as DESI, the William Herschel Telescope Enhanced Area Velocity Explorer (WEAVE), the Subaru Prime Focus Spectrograph (PFS), and the Chinese Space Station Telescope (CSST), will significantly increase the number of M31 objects with LOS velocity measurements. In addition, the forthcoming {\it Gaia} DR4 is expected to provide much more precise PM measurements. Together, these advances will enable more accurate constraints on the velocity anisotropy parameter $\beta$ and other kinematic properties of M31, offering deeper insights into M31's dynamical state and mass distribution.

\section*{Acknowledgements}
We are grateful to the referee for the constructive comments, particularly regarding the discussion of non-equilibrium effects and the inclusion of the Einasto profile, which have helped us improve the clarity and completeness of this work. We are grateful to Prof. Juntai Shen and Prof. Fangzhou Jiang for their invaluable comments and insightful discussions. This work acknowledges the supports from National Key R\&D Programme of China (Grant Nos. 2024YFA1611903 and 2025YFF0510603), the National Science Foundation of China (NSFC Grant No. 12422303, 12090040 and 12090044), the Young Scholar Program of Beijing Academy of Science and Technology (24CE-YS-08) and Beijing Natural Science Foundation (No. 1242016)

This work made use of the data from LAMOST (Large Sky Area Multi-Object Fiber Spectroscopic Telescope, also known as the Guoshoujing Telescope) (https://cstr.cn/31118.02.LAMOST). LAMOST is a Chinese national mega-science facility, operated by National Astronomical Observatories, Chinese Academy of Sciences.

\clearpage
\appendix

\section{Priors in the Hierarchical Bayesian Modeling}
\label{app:GMM_priors}

Here we list the priors adopted in the hierarchical Bayesian modeling of the dynamically hot halo velocity dispersion measurements described in Section~\ref{subsec:halo vel disp}. Table~\ref{tab:GMM_priors} lists the priors for Regions 2, 4, and 5, while Table~\ref{tab:GMM_priors_36} lists those for Regions 3 and 6.

\renewcommand{\thetable}{A\arabic{table}}
\setcounter{table}{0}

\renewcommand{\arraystretch}{1.2}

\begin{table*}
\centering
\caption{
Free parameters and priors used in the hierarchical Bayesian modeling for Regions 2, 4, and 5.}
\label{tab:GMM_priors}
\resizebox{\textwidth}{!}{ 
\begin{tabular}{cccccccc}
\hline
\hline
$R_{p}$ (kpc) & Region (\#) &  Parameter$^1$ & Allowed Range$^2$ & & Prior$^3$ & & Additional Constraints \\
\cline{5-7}
& & & & Form & Mean & Deviation & \\
\hline
\multirow{23}{*}{\makecell[c]{$6.6 < R_p \leq 17.8$}}
   & \multirow{2}{*}{Shared parameter} 
    & $\mu_{\rm LOS}^{\rm Halo}$ & [$-$50, 0] & Normal & $-$18 & 5 & --\\
    & & $\sigma_{\rm LOS}^{\rm Halo}$ & [5, 300] & Uniform & -- & -- & -- \\
\cline{2-8}
  & \multirow{7}{*}{2 (614)} 
  & $\mu_{\rm LOS}^{\rm KCC1}$        & [20, 120] & Uniform & -- & -- & -- \\
  &  & $\sigma_{\rm LOS}^{\rm KCC1}$  & [0, 80] & Uniform & -- & -- & $\sigma_{\rm LOS}^{\rm KCC1}<\sigma_{\rm LOS}^{\rm Halo}$ \\
  &  & $\mu_{\rm LOS}^{\rm KCC2}$     & [150, 300]  & Uniform & -- & -- & -- \\
  &  & $\sigma_{\rm LOS}^{\rm KCC2}$  & [0, 80]  & Uniform & -- & -- & $\sigma_{\rm LOS}^{\rm KCC2}<\sigma_{\rm LOS}^{\rm Halo}$ \\
  &  & $\mu_{\rm LOS}^{\rm Fgd}$      & [250, 350] & Normal & 300 & 50 & -- \\
  &  & $\sigma_{\rm LOS}^{\rm Fgd}$   & [0, 40] & Normal & 30 & 10 & -- \\
  &  & $f_{\rm Halo}$, $f_{\rm KCC1}$, $f_{\rm KCC2}, f_{\rm Fgd}$  & [0, 1] & Uniform & -- & -- & $f_{\rm Fgd}\leq0.1$; $f_{\rm Halo}+ f_{\rm KCC1}+f_{\rm KCC2}+f_{\rm Fgd}=1$ \\
\cline{2-8}
  & \multirow{7}{*}{4 (496)} 
  & $\mu_{\rm LOS}^{\rm KCC1}$        & [50, 150] & Uniform & -- & -- & -- \\
  &  & $\sigma_{\rm LOS}^{\rm KCC1}$  & [0, 80] & Uniform & -- & -- & $\sigma_{\rm LOS}^{\rm KCC1}<\sigma_{\rm LOS}^{\rm Halo}$ \\
  &  & $\mu_{\rm LOS}^{\rm KCC2}$     & [$-$300, $-$100]  & Uniform & -- & -- & -- \\
  &  & $\sigma_{\rm LOS}^{\rm KCC2}$  & [0, 80]  & Uniform & -- & -- & $\sigma_{\rm LOS}^{\rm KCC2}<\sigma_{\rm LOS}^{\rm Halo}$ \\
  &  & $\mu_{\rm LOS}^{\rm Fgd}$      & [250, 350] & Normal & 300 & 50 & -- \\
  &  & $\sigma_{\rm LOS}^{\rm Fgd}$   & [0, 40] & Normal & 30 & 10 & -- \\
  &  & $f_{\rm Halo}$, $f_{\rm KCC1}$, $f_{\rm KCC2}, f_{\rm Fgd}$  & [0, 1] & Uniform & -- & -- & $f_{\rm Fgd}\leq0.1$; $f_{\rm Halo}+ f_{\rm KCC1}+f_{\rm KCC2}+f_{\rm Fgd}=1$ \\
\cline{2-8}
  & \multirow{7}{*}{5 (260)} 
  & $\mu_{\rm LOS}^{\rm KCC1}$        & [50, 200] & Uniform & -- & -- & -- \\
  &  & $\sigma_{\rm LOS}^{\rm KCC1}$  & [0, 80] & Uniform & -- & -- & $\sigma_{\rm LOS}^{\rm KCC1}<\sigma_{\rm LOS}^{\rm Halo}$ \\
  &  & $\mu_{\rm LOS}^{\rm KCC2}$     & [$-$250, $-$50]  & Uniform & -- & -- & -- \\
  &  & $\sigma_{\rm LOS}^{\rm KCC2}$  & [0, 80]  & Uniform & -- & -- & $\sigma_{\rm LOS}^{\rm KCC2}<\sigma_{\rm LOS}^{\rm Halo}$ \\
  &  & $\mu_{\rm LOS}^{\rm Fgd}$      & [250, 350] & Normal & 300 & 50 & -- \\
  &  & $\sigma_{\rm LOS}^{\rm Fgd}$   & [0, 40] & Normal & 30 & 10 & -- \\
  &  & $f_{\rm Halo}$, $f_{\rm KCC1}$, $f_{\rm KCC2}, f_{\rm Fgd}$  & [0, 1] & Uniform & -- & -- & $f_{\rm Fgd}\leq0.1$; $f_{\rm Halo}+ f_{\rm KCC1}+f_{\rm KCC2}+f_{\rm Fgd}=1$ \\
\hline
\multirow{16}{*}{\makecell[c]{$17.8 < R_p \leq 27.4$}}
   & \multirow{2}{*}{Shared parameter} 
    & $\mu_{\rm LOS}^{\rm Halo}$ & [$-$50, 0] & Normal & $-$18 & 5 & --\\
    & & $\sigma_{\rm LOS}^{\rm Halo}$ & [5, 300] & Uniform & -- & -- & -- \\
\cline{2-8}
  & \multirow{7}{*}{2 (487)} 
  & $\mu_{\rm LOS}^{\rm KCC1}$        & [50, 150] & Uniform & -- & -- & -- \\
  &  & $\sigma_{\rm LOS}^{\rm KCC1}$  & [0, 80] & Uniform & -- & -- & $\sigma_{\rm LOS}^{\rm KCC1}<\sigma_{\rm LOS}^{\rm Halo}$ \\
  &  & $\mu_{\rm LOS}^{\rm KCC2}$     & [$-$30, 10]  & Normal & $-10$ & 10 & -- \\
  &  & $\sigma_{\rm LOS}^{\rm KCC2}$  & [0, 65]  & Normal & 55 & 10 & $\sigma_{\rm LOS}^{\rm KCC2}<\sigma_{\rm LOS}^{\rm Halo}$ \\
  &  & $\mu_{\rm LOS}^{\rm Fgd}$      & [250, 350] & Normal & 300 & 50 & -- \\
  &  & $\sigma_{\rm LOS}^{\rm Fgd}$   & [0, 40] & Normal & 30 & 10 & -- \\
  &  & $f_{\rm Halo}$, $f_{\rm KCC1}$, $f_{\rm KCC2}$, $f_{\rm Fgd}$  & [0, 1] & Uniform & -- & -- & \makecell[c]{$f_{\rm Fgd}\leq0.1$; $f_{\rm Halo}\leq0.3$; \\ $f_{\rm Halo}+ f_{\rm KCC1}+f_{\rm KCC2}+f_{\rm Fgd}=1$} \\
\cline{2-8}
  & \multirow{7}{*}{4 (388)} 
  & $\mu_{\rm LOS}^{\rm KCC1}$        & [0, 120] & Uniform & -- & -- & -- \\
  &  & $\sigma_{\rm LOS}^{\rm KCC1}$  & [0, 80] & Uniform & -- & -- & $\sigma_{\rm LOS}^{\rm KCC1}<\sigma_{\rm LOS}^{\rm Halo}$ \\
  &  & $\mu_{\rm LOS}^{\rm KCC2}$     & [$-$200, $-$50]  & Uniform & -- & -- & -- \\
  &  & $\sigma_{\rm LOS}^{\rm KCC2}$  & [0, 80]  & Uniform & -- & -- & $\sigma_{\rm LOS}^{\rm KCC2}<\sigma_{\rm LOS}^{\rm Halo}$ \\
  &  & $\mu_{\rm LOS}^{\rm Fgd}$      & [250, 350] & Normal & 300 & 50 & -- \\
  &  & $\sigma_{\rm LOS}^{\rm Fgd}$   & [0, 40] & Normal & 30 & 10 & -- \\
  &  & $f_{\rm Halo}$, $f_{\rm KCC1}$, $f_{\rm KCC2}, f_{\rm Fgd}$  & [0, 1] & Uniform & -- & -- & $f_{\rm Fgd}\leq0.1$; $f_{\rm Halo}+ f_{\rm KCC1}+f_{\rm KCC2}+f_{\rm Fgd}=1$ \\
\hline
\multirow{9}{*}{\makecell[c]{$27.4 < R_p \leq 39.6$}}
  & \multirow{9}{*}{2 (291)} 
    & $\mu_{\rm LOS}^{\rm Halo}$ & [$-$50, 0] & Normal & $-$18 & 5 & --\\
    & & $\sigma_{\rm LOS}^{\rm Halo}$ & [5, 300] & Uniform & -- & -- & -- \\
    & & $\mu_{\rm LOS}^{\rm KCC1}$        & [$-$20, 20] & Normal & 0 & 10 & -- \\
    & & $\sigma_{\rm LOS}^{\rm KCC1}$  & [0, 65] & Normal & 45 & 10 & $\sigma_{\rm LOS}^{\rm KCC1}<\sigma_{\rm LOS}^{\rm Halo}$ \\
    & & $\mu_{\rm LOS}^{\rm Fgd}$      & [250, 350] & Normal & 300 & 50 & -- \\
    & & $\sigma_{\rm LOS}^{\rm Fgd}$   & [0, 40] & Normal & 30 & 10 & -- \\
    & & $f_{\rm Halo}$, $f_{\rm KCC1}$, $f_{\rm Fgd}$  & [0, 1] & Uniform & -- & -- & \makecell[c]{$f_{\rm Fgd}\leq0.1$; $f_{\rm Halo}\leq0.6$; \\ $f_{\rm Halo}+ f_{\rm KCC1}+f_{\rm Fgd}=1$} \\
\hline
\end{tabular}
}
\begin{flushleft}
\footnotesize
\textbf{Notes.} \\
(1) The parameters used to describe each component include the mean line-of-sight velocity ($\mu_{\rm LOS}$; in $\rm km\,s^{-1}$), the velocity dispersion ($\sigma_{\rm LOS}$; in $\rm km\,s^{-1}$), and the corresponding normalized fractional contribution ($f$).\\
(2) The velocities are the peculiar LOS velocity transformed in Section~\ref{subsec:Velocity Trans}.\\
(3) The mean and standard deviation are applied only to normal priors.
\end{flushleft}
\end{table*}

\renewcommand{\arraystretch}{1.2}
\begin{table*}
\centering
\caption{
Free parameters and priors used in the hierarchical Bayesian modeling for Regions 3 and 6.}
\label{tab:GMM_priors_36}
\resizebox{\textwidth}{!}{ 
\begin{tabular}{cccccccc}
\hline
\hline
$R_{p}$ (kpc) & Region (\#) &  Parameter$^1$ & Allowed Range$^2$ & & Prior$^3$ & & Additional Constraints \\
\cline{5-7}
& & & & Form & Mean & Deviation & \\
\hline
\multirow{11}{*}{\makecell[c]{$15.5 < R_p \leq 36.9$}}
  & \multirow{11}{*}{3 (345)} 
  & $\mu_{\rm LOS}^{\rm Halo}$ & [$-$50, 0] & Normal & $-$18 & 5 & --\\
  & & $\sigma_{\rm LOS}^{\rm Halo}$ & [5, 300] & Uniform & -- & -- & -- \\
  & & $\mu_{\rm LOS}^{\rm KCC1}$        & [$-200$, 0] & Uniform & -- & -- & -- \\
  & & $\sigma_{\rm LOS}^{\rm KCC1}$  & [0, 80] & Uniform & -- & -- & $\sigma_{\rm LOS}^{\rm KCC1}<\sigma_{\rm LOS}^{\rm Halo}$ \\
  & & $\mu_{\rm LOS}^{\rm KCC2}$     & [100, 200]  & Uniform & -- & -- & -- \\
  & & $\sigma_{\rm LOS}^{\rm KCC2}$  & [0, 80]  & Uniform & -- & -- & $\sigma_{\rm LOS}^{\rm KCC2}<\sigma_{\rm LOS}^{\rm Halo}$ \\
  & & $\mu_{\rm LOS}^{\rm KCC3}$     & [$-30$, 30]  & Uniform & -- & -- & -- \\
  & & $\sigma_{\rm LOS}^{\rm KCC3}$  & [0, 30]  & Uniform & -- & -- & $\sigma_{\rm LOS}^{\rm KCC3}<\sigma_{\rm LOS}^{\rm Halo}$ \\
  & & $\mu_{\rm LOS}^{\rm Fgd}$      & [250, 350] & Normal & 300 & 50 & -- \\
  & & $\sigma_{\rm LOS}^{\rm Fgd}$   & [0, 40] & Normal & 30 & 10 & -- \\
  & & $f_{\rm Halo}$, $f_{\rm KCC1}$, $f_{\rm KCC2}, f_{\rm KCC3}, f_{\rm Fgd}$  & [0, 1] & Uniform & -- & -- & $f_{\rm Fgd}\leq0.1$; $f_{\rm Halo}+ f_{\rm KCC1}+f_{\rm KCC2}+f_{\rm KCC3}+f_{\rm Fgd}=1$ \\
\hline
\multirow{7}{*}{\makecell[c]{$14.6 < R_p \leq 50.1$}}
  & \multirow{7}{*}{6 (383)} 
  & $\mu_{\rm LOS}^{\rm Halo}$ & [$-$50, 0] & Normal & $-$18 & 5 & --\\
  & & $\sigma_{\rm LOS}^{\rm Halo}$ & [5, 300] & Uniform & -- & -- & -- \\
  & & $\mu_{\rm LOS}^{\rm KCC1}$        & [$-300$, $-100$] & Uniform & -- & -- & -- \\
  & & $\sigma_{\rm LOS}^{\rm KCC1}$  & [0, 80] & Uniform & -- & -- & $\sigma_{\rm LOS}^{\rm KCC1}<\sigma_{\rm LOS}^{\rm Halo}$ \\
  & & $\mu_{\rm LOS}^{\rm Fgd}$      & [250, 350] & Normal & 300 & 50 & -- \\
  & & $\sigma_{\rm LOS}^{\rm Fgd}$   & [0, 40] & Normal & 30 & 10 & -- \\
  & & $f_{\rm Halo}$, $f_{\rm KCC1}$, $f_{\rm Fgd}$  & [0, 1] & Uniform & -- & -- & $f_{\rm Fgd}\leq0.1$; $f_{\rm Halo}+ f_{\rm KCC1}+f_{\rm Fgd}=1$ \\
\hline
\end{tabular}
}
\begin{flushleft}
\footnotesize
\textbf{Notes.} \\
(1) The parameters used to describe each component include the mean line-of-sight velocity ($\mu_{\rm LOS}$; in $\rm km\,s^{-1}$), the velocity dispersion ($\sigma_{\rm LOS}$; in $\rm km\,s^{-1}$), and the corresponding normalized fractional contribution ($f$).\\
(2) The velocities are the peculiar LOS velocity transformed in Section~\ref{subsec:Velocity Trans}.\\
(3) The mean and standard deviation are applied only to normal priors.
\end{flushleft}
\end{table*}

\newpage
\bibliography{ref}
\bibliographystyle{aasjournal}

\end{document}